%% file: main.tex
\documentclass[twocolumn]{article}

\usepackage{graphicx} 

\usepackage{t1enc}
\usepackage[utf8]{inputenc}
\usepackage{wrapfig}
\usepackage{todonotes}
\usepackage{array,graphicx,multicol}
\usepackage{booktabs}
\usepackage{pifont}
\usepackage{xspace}
\usepackage{hyperref}
\usepackage[inline]{enumitem}
\usepackage{subfig}
\usepackage{rotating}
\usepackage{cite}
\newcommand*\OK{\ding{51}}

\newcommand{\etal}{\emph{et al.}}
\newcommand{\eg}{\emph{e.g.,}\xspace}

\title{The State of the Art in Multilayer Network Visualization}

\author{Fintan McGee$^{1}$,
        Mohammd Ghoniem$^{1}$\\
        Guy Melan{\c{c}}on$^{2}$,
        Benoit Otjacques$^{1}$,
        and Bruno Pinaud$^{2}$ 
        \newline
        \\
         $^1$Luxembourg Institute of Science and Technology (LIST),\\
         firstname.lastname@list.lu\\
         $^2$University of Bordeaux, LaBRI UMR CNRS 5800, France\\
         firstname.lastname@u-bordeaux.fr
       }

\begin{document}
\newcommand{\FMrevised}[1]{\textcolor{black}{#1}}
\newcommand{\MGrevised}[1]{\textcolor{black}{#1}}
\newcommand{\BPrevised}[1]{\textcolor{black}{#1}}
\newcommand{\GMrevised}[1]{\textcolor{black}{#1}}
\newcommand{\BOrevised}[1]{\textcolor{black}{#1}}
\newcommand{\BOcomment}[1]{\todo[color=black]{#1}}


\maketitle

\input{1_Abstract_Introduction.tex}
\input{2_Multilayer_Graph_Definition.tex}
\input{3_Survey.tex}
\input{4_Discussion.tex}

\input{5_Conclusions_End.tex}


\bibliographystyle{plain}

\bibliography{MultilayerBibliography}


\end{document}

%% file: 1_Abstract_Introduction.tex
\begin{abstract}
Modelling relationships between entities in real-world systems with a simple graph is a standard approach. However, reality is better embraced as several interdependent subsystems (or layers).
	Recently the concept of a multilayer network model has emerged from the field of complex systems.
	This model can be applied to a wide range of real-world datasets.
	Examples of multilayer networks can be found in the domains of life sciences, 
sociology, digital humanities and more.
	Within the domain of graph visualization there are many systems which visualize datasets having many characteristics of multilayer graphs.
This report provides a state of the art and a structured analysis of contemporary multilayer network visualization, not only for researchers in visualization, but also for those who aim to visualize multilayer networks in the domain of complex systems, as well as those developing systems across application domains. We have explored the visualization literature to survey visualization techniques suitable for multilayer graph visualization, as well as tools, tasks, and analytic techniques from within application domains.
This report also identifies the outstanding challenges for multilayer graph visualization and suggests future research directions for addressing them.

%

\end{abstract}  
\section{Introduction}
\label{sect:intro}
Simple graphs are often used to model relationships between entities in real-world systems. 
This approach may however be an oversimplification of a much more complex reality better embraced as several interdependent subsystems (or layers), which motivated the development of the \emph{complex networks} field~\cite{gao2012networks, KENETT2015}.
The concept of a multilayer network~\cite{MultilayerNetworks} builds on and encompasses many existing network definitions across many fields, some of which are much older, \eg from the domain of sociology~\cite{moreno1953,verbrugge1979multiplexity,Burt1985}.
  

As an introductory illustrative example, consider a person's social networks. 
People frequently use more than one social network platform, \eg Facebook for their personal social network or LinkedIn for their professional. Offline, "real life", social networks could also be considered, again with relations being either personal or professional.
These networks can be considered independent, however they
can also be considered as layers in a multilayer graph.
The networks overlap  as some people may be present across layers.
Layers are in this case characterised by relationship type (either online/offline and personal/professional).
A significant change in one network may implicitly correlate with or cause changes in another. 
For example, a change of employer will cause changes in both offline and online professional networks but in a different manner for each, and may cause slower, more gradual, changes in the personal offline/online social networks.
To answer some questions, it may be necessary to also include employers or companies as entities of the network. 
This makes it possible to model explicitly person-company relationships, as well as person-person and company-company relationships.
In this case, layers may be characterised by entity type (either person or company).
Other definitions of layers are also possible as illustrated in Section~\ref{sect:origin}.

Examples of multilayer networks can be found in the domains of biology (the so-called ``omics'' layers), epidemiology~\cite{wang2012epidemics,saumell2012epidemic, pastor2015epidemic}, sociology (in a broad sense, including fields such as criminology, for instance) \cite{Burt1985,Lazega1999,Geard2007,freire2010ManyNets,Ghani:2013fo,crnovrsanin2014visualization,Bright2015,Dickison2016}, digital humanities~\cite{mcgee2016towards,dunne2012graphtrail, Sluban2016Temporal}, civil infrastructure~\cite{cardillo2013emergence,ducruet2017Maritime,derrible2017complexity} and more.
Multilayer networks have been explicitly recognised as promising for biological analysis~\cite{gosak2017}. We give more details in Section~\ref{subsec:domains}.

In the area of network visualization many systems visualize datasets having many characteristics of multilayer networks, albeit under a different title.
Multi-label, multi-edge, multi-relational, multiplex~\cite{renoust2015detangler,cardillo2013emergence}, heterogeneous~\cite{dunne2012graphtrail,schreiber2014heterogeneous}, and multimodal\cite{Ghani:2013fo,heath2009multimodal}, multiple edge set networks\cite{crnovrsanin2014visualization}, interdependent networks~\cite{gao2012networks}, interconnected networks\cite{saumell2012epidemic} and networks of networks\cite{KENETT2015} are amongst the many names given to various types of data that are encapsulated by the Multilayer Networks definition of Kivel{\"a} \emph{et al.}~\cite{MultilayerNetworks}.


Recently initial steps have been made towards consolidating the work on visualization of multilayer networks from domains outside of the information visualization field, see \emph{MuxVis}~\cite{dedominico2015MuxViz} from the domain of complex systems, or from the domain of social networks~\cite{dickison_magnani_rossi_2016}, based on the complex systems paper of Rossi and Magnani~\cite{rossi2015}.
However, to date there has been no survey quantifying and consolidating the state of the art of visualization of multilayer networks, both within the field of information visualization and across application domains. 

The goal of this survey is to reconcile the many visualization approaches from the information visualization field and the application domains and group them together as a consistent set of techniques to support the increasing demand for the visualization of multilayer networks.
The final contribution of this work consists in identifying the key challenges outstanding in the field, and providing a road map for future research developments on the topic.

This report is structured as follows: 
Section~\ref{sect:origin} presents the defining concepts underlying multilayer graph models, and points out the main differences they have with other related network models. The rest of the section briefly describes the application domains in which multilayer graphs are encountered.
The description of the methodology followed is presented in
Section~\ref{sect:survey_methodo} followed in Section~\ref{sect:thesurvey} by the survey itself. It provides a structured account of relevant tasks, visualization and interaction techniques pertaining to multilayer network analysis.  
In Section~\ref{sect:discussion} we reflect on the state of the art in multilayer network visualization, and point out open challenges and opportunities that lie ahead of the information visualization research community.
We finish this paper in Section~\ref{sect:conclusion} with concluding remarks and a roadmap for future contributions to the topic of multilayer networks visualization.


%% file: 2_Multilayer_Graph_Definition.tex
\section{Multilayer Networks and Related Concepts}\label{sect:origin}
The notion of many relationships between individuals, often called \emph{multiplex} relationships,
is seminal in sociology and one could argue that it already was present in the sociograms introduced by Moreno~\cite{moreno1953}. The notion is central in the work of Burt and Sch{\o}tt~\cite{Burt1985} where the challenge is to somehow simplify multiplex relationships, consolidate and substitute them for relationships involving a smaller number of relation types to ease the analysis of the network. More recently, the concept of a multilayer network has emerged from the Complex Networks area, a subdomain of the field of complex systems, and is a fertile ground for novel visualization research. 

\subsection{Defining concepts}
\label{subsec:layer_related_concepts}

It is important to emphasise that layers do not reduce to some operational apparatus. The concept goes far beyond a simple intent to capture data heterogeneity. While it is true this notion is most of the time embodied as nodes and edges of a network being of different ``types'', its roots lie deeply in sociology~\cite{Burt1985,Lazega1999,Geard2007}. This notion is used to form questions and hypotheses, where layers can be considered as innermost, intermediate or outer \cite{lin2008}. For instance, \BPrevised{Dunbar \etal}~\cite{DUNBAR2015} consider networks similar to our introductory example, and examine to what extent online and offline layers in personal networks overlap.

While innermost and outermost layers are well established notions in sociology, the modeller is free to be ``creative'' when deciding what constitutes a layer (\emph{dixit} Kivel\"a \emph{et al.}~\cite{MultilayerNetworks}). That is, the notion of a layer in a network emerges from and belongs to the domain under investigation.
Consequently, when discussing the notion of layer, it is important to distinguish the sociological network from the mathematical network used to describe it. The mathematical network -- a graph -- is but an artefact through which we may hope to observe and ultimately characterise a phenomenon occurring on the sociological network. The definition of a layer is thus a characteristic of the multilayer system as a whole, defined either by a physical reality or the system being modelled. The notion of a layer naturally occurs when describing tasks performed by analysts; it can be mobilised to form exploration or browsing strategies (see Section \ref{subsec:task_analysis}).

\paragraph*{Formal Definition.}
A standard graph is often described by a tuple $G=(V,E)$ where $V$ defines a set of vertices and $E$ defines a set of edges (vertex pairs), such that $E \subseteq V \times V$. An intuitive definition of a multilayer network first consists in specifying which layers nodes belong to. Because we allow a node $v \in V$ to be part of some layers and not to others, we may consider `multilayer graph' nodes as pairs $V_M \subseteq V \times L$ where $L$ is the set of considered layers. Edges $E_M \subseteq V_M \times V_M$ then connect pairs $(v, l), (v', l')$. An edge is often said to be \emph{intra} or \emph{inter}-layer depending on whether $l = l'$ or $l \not = l'$.

Going back to the example where people use different social network platforms, we would have $L = \{l, l', l'', \ldots\}$ where $l = $ Facebook friends, $l' =$ LinkedIn connections, $l'' =$ ``real life'' family-friends-acquaintances, etc.

\subsection{Aspects}
Kivel{\"a} \emph{et al.} also define what they call \emph{aspects} as a way to characterise a set of elementary layers relating to some concepts. An example would be:
\begin{itemize}
\item aspect $L_1$ capturing interaction between people in the context of their participation to events (\BPrevised{\eg conferences~\cite{Atzmueller2012}}), with $l_1$ for interaction during InfoVis, $l_2$ for interaction during EuroVis, etc.);
\item aspect $L_2$ capturing co-authorship around themes (an example we borrow from \BPrevised{Renoust \etal}~\cite{renoust2015detangler}), with $l_i$ for co-authorship associated with some keyword $k_i$;
\item aspect $L_3$ capturing project partnership, with layers $l_i$ associated with specific programs, for example~\cite{Ghani:2013fo};
\item and so forth.
\end{itemize}

Aspects can also be used as an artefact to deal with time or geographical position.

\begin{figure}[!ht]
	\centering
		\includegraphics[width=\columnwidth]{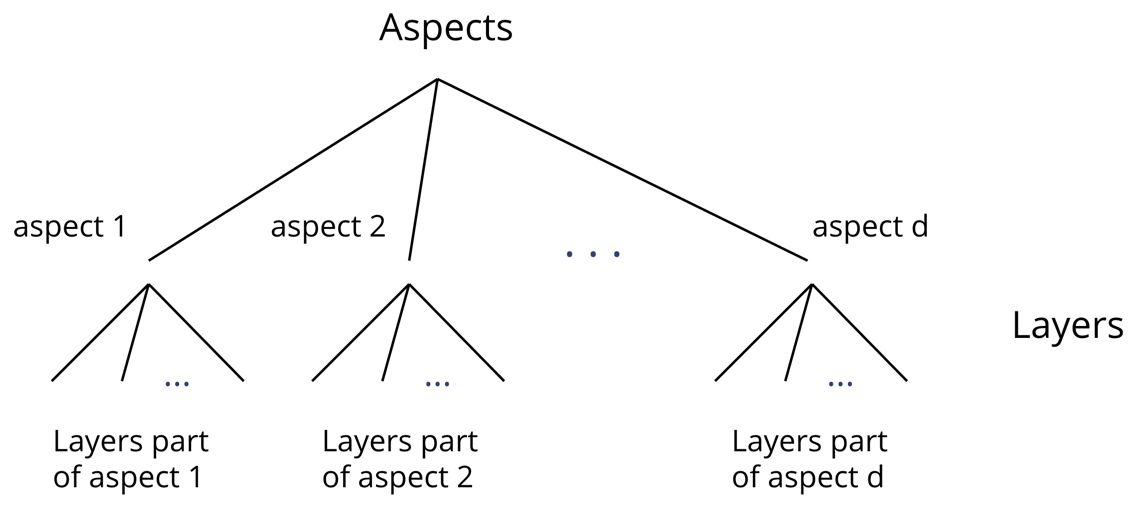}
	\caption{Aspects can be seen as groups of layers of different types. Nodes do not necessarily appear on all layers, but they necessarily appear on at least one layer of each aspect.}
	\label{fig:aspect_illustration}
\end{figure}

Aspects can be captured by extending the previous definition, as proposed Kivel{\"a} \emph{et al.}:

Given any number $d$ of aspects, ${\textbf{L}} = \{L_1, L_2, \ldots, L_d\}$, a multilayer network corresponds to a quadruple  $M = (V_M,E_M,V,\textbf{L})$, where each aspect $L_a$ is a \emph{set of elementary layers} and $V_M \subseteq V \times L_1 \times \dots L_d$. That is, while nodes do not necessarily appear on all elementary layers, they necessarily appear on at least one layer of each aspect. The set of edges of $M$ simply is $E_M \subseteq V_M \times V_M$ (see Figure~\ref{fig:aspect_illustration}). 

Kivel{\"a} \BPrevised{\etal} chose the term carefully, to avoid using a term that may be unclear depending on the reader's domain.
While the term dimension, in its literal meaning, may lend itself to the concept of defining a characteristic, \emph{aspect} has been chosen due to the use of the term dimension as jargon in different domains.

\begin{figure}[!ht]
	\centering
		\includegraphics[width=\columnwidth,keepaspectratio]{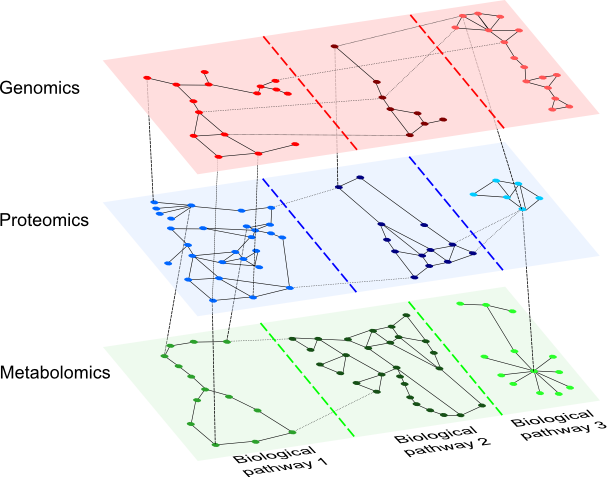}
	\caption{A purely illustrative example of multilayer data in the context of biology. The layer can be described by the type of data as a first aspect (genomic, proteomic, or metabolomic), and biological pathway being represented as second aspect.}
	\label{fig:layer_illustration}
\end{figure}

Another example lies in the domain of biology (described further in Section~\ref{subsec:domains}). One aspect is the type of data, such as genomic, \BPrevised{metabolomic} or proteomic. Another aspect might be the species, or different biological pathways, as illustrated in Figure~\ref{fig:layer_illustration}. If the biological data contains time information, that may also be considered an aspect. While multiple aspects are a possibility for multilayer network data sets, it is not a requirement. A multilayer data set may be defined by a single aspect, which categorises multiple layers.
See Table~\ref{tab:AspectExamples} for a sample list of aspects and layers extracted from the literature surveyed as part of this report.
Kivel\"{a} \emph{et al.}~\cite{MultilayerNetworks} provide further examples in their extensive list of multiplex datasets and their associated layers.

\begin{table*}[!ht]
	\centering
		\begin{tabular}{|l|l|l|l|}
		\hline	
		Aspect Description & Layer Definition & Source Paper &Source Paper Domain \\ \hline
		Social entity type & People, societies / organisations & \cite{renoust2015detangler} & Information visualization \\
                Social relationship type & Friendship, aggression & \cite{crnovrsanin2014visualization} & Social networks\\
		Word relationship &	Hyponym, homonym	 & \cite{Hascoet:2012:IGM:2254556.2254654} & Information visualization\\
		Year of publication & [1974...2004] & \cite{Hascoet:2012:IGM:2254556.2254654} & Information visualization\\
        \shortstack[c]{Infrastructure\\ connection type} & Air connection, train connection & \cite{Halu2014} &Physics\\
                Transport mode & air, rail, ferry, coach & \cite{Barthelemey15} & Scientific data (Transportation)\\
        ``Omics'' Entity type & Gene, protein, protein structure & \cite{pavlopoulos2008arena3d} & Biology\\
        Historical correspondences &\shortstack[c]{Letter, letter sender,\\ letter receiver, cited book} & \cite{vanVugt2017letters}& Historical network research \\
        Building layout & Arrangement of house spaces & \cite{SLUSARCZYK201795} & Robot Control Algorithms\\
		\hline
		\end{tabular}
	\caption{Examples of aspects and layers, extracted from papers covered by this survey. }
	\label{tab:AspectExamples}
\end{table*}

Incidentally, \BPrevised{Wehmuth \etal}~\cite{WEHMUTH201650} propose an alternative definition they call MultiAspect graphs where they formally define what can be considered as an aspect. Unsurprisingly, they also form a network where nodes are defined using Cartesian products collecting multiple values into a single entity.
The authors describe MultiAspect graphs as forming a generalisation of Kivel{\"a} \emph{et al.}'s multilayer network. Reconciling these different approaches is beyond the scope of this paper. Well developed examples are certainly needed to uncover the full applicative potential of MultiAspect graphs. 

\subsection{Related Graph Models}

\begin{figure*}[!ht]
  \centering
  \subfloat[\FMrevised{An n-partite graph (n = 3).}\label{subfig-1:n}] {
   \includegraphics[width=0.5\columnwidth]{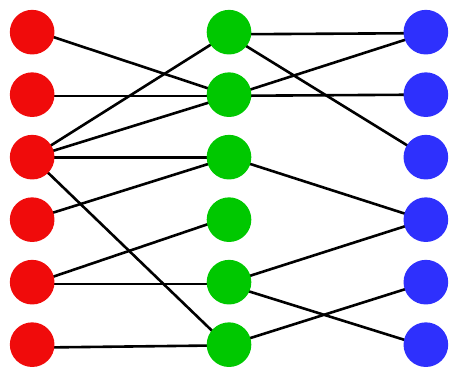}
   }
   \hfill
    \subfloat[\FMrevised{A Multivariate graph, where each data node contains multiple attributes.}\label{subfig-1:mv}] {
   \includegraphics[width=0.7\columnwidth]{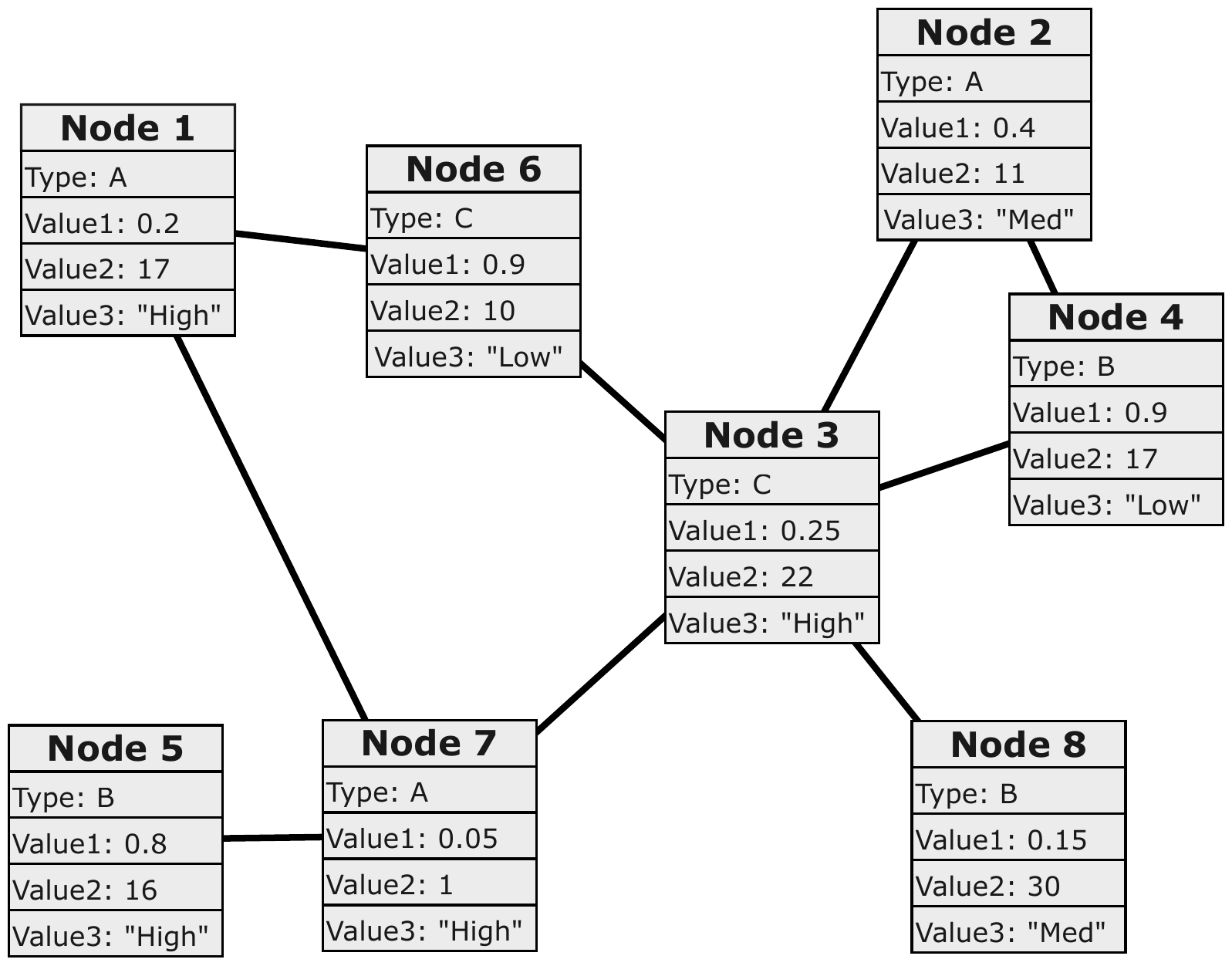}
   }
   \hfill
   \subfloat[\FMrevised{A dynamic graph, with 4 time slices, where structure changes over time.}\label{subfig-1:d}] {
    \includegraphics[width=0.6\columnwidth]{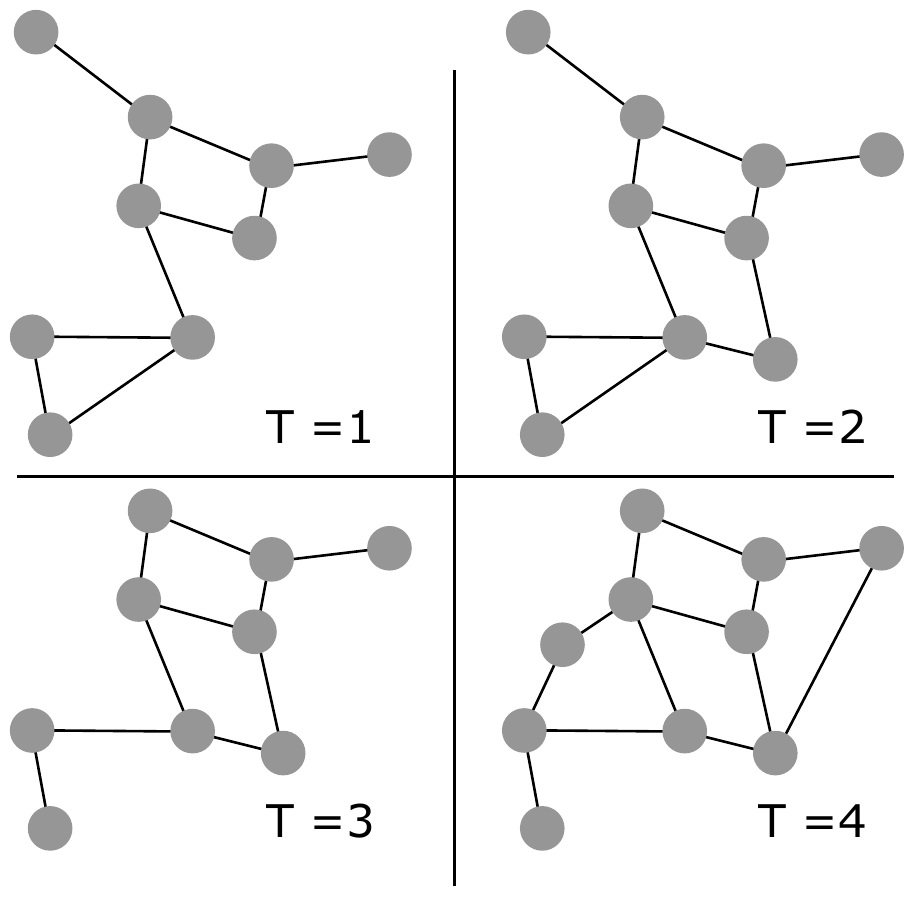}
    }
  \caption{\FMrevised{Illustrative examples of related graph models: Each of the three nodes types (*indicated by colour) of the n-partite graph could define a layer within a multilayer network, in this case all edges would be between layers. For a multivariate graph, node attributes could be used do divide the network into layers. Defining layers by node type in this example would result in three layers, although that may not make sense for the system being modelled, as there would be no edges within the layers of nodes of type B and C. For a dynamics graph characterized by time slices, each time slice can be intuitively understood as a layer. Further insight could be gained by by the use of an additional aspect to define layers.}}
  \label{fig:relatedModels}
\end{figure*}
Below, we review related graph models \FMrevised{(see also Figure \ref{fig:relatedModels})} and their differences or resemblances to multilayer networks.
\subsubsection{N-partite Graphs}


Recall that a bipartite graph is made of two disjoint sets of vertices so that no two vertices belonging to the same set are connected.
Bipartite graphs can be considered as a case of multilayer networks with 2 layers and only interlayer edges. The two mode (\emph{i.e.}, node type) nature of bipartite graphs result in analytics that are different to those of single mode graphs~\cite{borgatti1997network}.
Bipartite graph concepts are sometimes extended into n-partite graphs, \FMrevised{as seen in our example in figure \ref{subfig-1:n}}, although in practice many of the 2 mode restrictions associated with bipartite graph are not fully retained. 
In practice, systems which model bipartite cases and extensions of bipartite cases, such as the multimodal networks of \BPrevised{Ghani \etal}~\cite{Ghani:2013fo}, and the Academic network analysed by \BPrevised{Shi \etal}~\cite{shi2014Hierarchical}, can be considered instances of multilayer networks. In this case the authors also make use of bipartite analytics (\eg adapted centrality metrics) to better understand their network structure.

Bipartite networks can be reduced to single mode networks via projection on a mode. Such an operation may be used to also define a layer in a multilayer network, if the projection results in a layer that reflects the reality of the system being modelled.
\subsubsection{Multivariate Graphs}
\label{sec:multivariate}
Multivariate graphs~\cite{kerren2014introduction} are those in which nodes or edges carry attributes or properties.
As described by Schreiber \emph{et al.}~\cite{schreiber2014heterogeneous}, there is a relationship between multivariate graphs and multilayer graphs. 
Some variables or attributes in a multivariate dataset often serve the purpose of distinguishing nodes and edges that belong to different layers, \eg the type of social network platform in our initial example. 
There are also multivariate visualization applications such as that of Pretorius and van Wijk~\cite{pretorius2008}, that define their graph as having discrete sets, which can be considered analogous to defining layers.
However, in the majority of cases research into multivariate visualization lacks the \emph{a priori} definition of a layer defined by a physical or conceptual reality related to the system being modelled.

In faceted datasets, multivariate data items are grouped in multiple orthogonal categories.
Originally used as an approach to search and browse large data stores and text corpora~\cite{smith2006facetmap,cao2010facetatlas}, later work extended the faceted approach to include relationship visualization~\cite{lee2009facetlens,zhao2013interactive}.
Datasets can have many different facets such as spatial and temporal frames of reference, or multiple values per data item and as such can be considered multifaceted.
Visualizations for multifaceted data are those which show more than one of these facets simultaneously (see Hadlak \etal~\cite{hadlak2015survey} for a survey of multifaceted graph visualization techniques).
Hadlak \emph{et al.} discuss primarily four common facets of network structure considered in network visualization, and their composition: partitions, attributes, time, and space.
These facets may be considered to be very similar to instances of  Kivel\"{a} \emph{et al.}'s aspects. However, they can be considered as different ways of exploring a single data set, (which is unsurprising given the origins of a faceted visualization).
The techniques described are still very useful for developing approaches for visualizing layers, particularly where the layer type matches the Hadlak \emph{et al.}'s selected  faceted categories. However faceted network visualization approaches do not meet all the needs for multilayer network visualization.
While multilayer networks may use notions similar to these facets to characterise layers, multilayer network visualization also focuses on the interactions between layers and the role of layers in the network as a whole. 

\subsubsection{Dynamic Graphs}
Dynamic graphs are graphs whose structure (nodes and edges) and/or associated attributes may change over time. Analysts are often interested in comparing the state of the network at different points in time.
Within the domain of complex networks Boccaletti \BPrevised{\etal}~\cite{boccaletti2014structure} consider the dynamics of multilayer networks, and in many cases time slices of a dynamic (or temporal) network are simply mapped to layers.
The notion of dynamic networks is also mentioned by Kivel\"{a} \emph{et al.}, who notes that they can be considered as a type of multilayer network. A set of dynamic time slices can be considered layers in an aspect representing time.
As multilayer networks can have multiple aspects, a temporal aspects might be just one of many.
In their report on dynamic network visualization Moody \emph{et al.}~\cite{moody2005dynamic} explain the importance of ``multiplicity'' in social networks, \emph{i.e.}, the overlap of types of relations.
In particular, they point out that linking relational timing to tie types allow to better investigate social dynamics.
A recent survey of dynamics graph visualization techniques  was provided by Beck \emph{et al.}~\cite{beck2014state}, but does not consider layers in any context other than a hierarchical graph.

\subsection{Application Domains and Data}\label{subsec:domains}
Across all of the application domains described in Section~\ref{sect:intro}, advances in sensors, scientific equipment, and technology mean that researchers have access to more data than ever.
This wealth of complex data is often best understood as a multilayer network model.
\paragraph*{Life Sciences:}
Within  biological network visualization there are many contexts in which a multilayer network approach may be beneficial~\cite{gosak2017}.
Biologists have access to more genomic, proteomic and metabolomic data, allowing for the construction of complex multilayer models of intricate biological processes. 
Interactions taking place within the genomic, proteomic and metabolomic levels can be modelled as individual networks, but interactions also occur between elements sitting in different omics levels within a larger biological system, where the aspect characterising the layer is the node type \cite{cottret2010}.
This corresponds to the strongly rising topic of systems/integrative biology, where the challenge consists in understanding the interplay and the cascade of effects taking place at the different levels of the biological system at hand\cite{gehlenborg2010visualization,kuo20133omics}.  
A prominent task for biologists analysing biological pathways consists in comparing a species-specific pathway to a reference pathway~\cite{Murray2017Taxonomy}, in this specific case species type can be considered a defining aspect for a layer.
Another task is to compare tissue-specific interaction networks to understand why certain tissues, \eg plant root tissues, synthesise certain molecules which are not found in other plant tissues. In this case tissue type is the defining aspect for a layer.

\paragraph*{Social Sciences:} 
Datasets within Social Network analysis frequently contain multiple types of edges (\eg looking at the different types of relationships between people, e.g., more recently~\cite{crnovrsanin2014visualization}, but also in much earlier work such as~\cite{Burt1985,Lazega1999}), or multiple types (or modes) of nodes \eg modelling a citation network containing researchers, institutions and publications~\cite{Ghani:2013fo}.
Within social sciences, there are also contexts in which many networks may be compared to one another. For example, examining social networks produced as a result of cell phone activity, as done by \BPrevised{Freire \etal}~\cite{freire2010ManyNets}.
The contemporary use of multiple online social networks provides a vast amount of data. This allows for complex social multilayer networks to be built, that may help sociologists gain deeper insight~\cite{Renoust2014}.

Other fields such as Food Microbiology, have adopted Social Network Analysis techniques, and applied them to understand problems such as the spread of disease. This can be seen in the work of Crabb \BPrevised{\etal}~\cite{crabb2017disease} to understand the spread of salmonella in a large poultry farming enterprise. Different networks are generated based on contact between different types of entities. From a multilayer perspective, contact between entities can be considered an aspect, with the entity types defining the different layers.

\paragraph*{Digital Humanities:} 
Within digital humanities fields, such as digital cultural heritage, archaeology and data journalism, many multilayer approaches~\cite{vanVugt2017letters,mcgee2016towards,dunne2012graphtrail,ren2018generating} can be found.
Digital access to source texts and natural language processing techniques such as Named-Entity Recognition and Topic Modelling allow for vast Digital Humanities datasets to be built~\cite{mcgee2016towards}. Co-occurrence relationships between people names, locations, organisations as well as other entities form a typical multilayer network whose analysis may reveal insightful interaction patterns.  

\paragraph*{Infrastructure:}
Modern vehicles often provide a wealth of information about modern transportation networks.
These networks can also be modelled as multilayer networks. 
For example, \BPrevised{Halu \etal}~\cite{Halu2014} models the air and rail transportation networks of India as layers in a multilayer network. \BPrevised{A paper by Gallotti and Berthelemy~\cite{Barthelemey15} is another example}.
The Internet and associated infrastructure provide vast amounts of data about themselves and can be modelled as multilayer networks, as done by Reis \emph{et al.}~\cite{reis2014}, who represent the power grid and the Internet as separate interdependent layers in a multilayer infrastructure network.
Recent work concerning Urban Infrastructure Systems highlight the necessity to adopt an integrated approach to urban planning taking into account the interplay between multiple networks like transportation networks, energy networks, telecommunication networks, water/wastewater networks~\cite{derrible2017complexity}. 
Some of the related objectives may be to reduce the cascading of failures across these networks~\cite{buldyrev2010catastrophic}, but also to develop an efficient repair strategy to restore services after disaster~\cite{shekhtman2016recent}.
The precise representation of buildings to support robot control algorithms is a related domain as seen in ~\cite{SLUSARCZYK201795}. In this work, the graph represents a layout of the floors of the building with their interconnections. A layer is a floor containing rooms. An edge represents a direct connection between two rooms. Interlayer connections modelled connections between floors. This kind of model reduces the number of data to be analysed by a robot.

The vast number of instances of complex datasets produced across all these examples demands a visual approach to help understand it,  and that approach will often be multilayer network visualization.

%% file: 3_Survey.tex
\section{Methodology Followed}
\label{sect:survey_methodo}
This section is about the structure of the survey which is built on a categorisation of the important features of multilayer network and how we select papers cited in the many domains we cover.

\subsection{Categorisation}
\label{sect:categorisation}
The categorisation of the most important features of multilayer network visualization that are to be considered for each paper is built in a manner consistent with Munzner's nested visualization design process model~\cite{Munzner2009}:

\paragraph*{Tasks and Analysis.} Multilayer systems that address new problems and domains may expose tasks that do not fit in existing task taxonomies, such as~\cite{lee2006task,pretorius2014tasks}. New analytics have been developed for multilayer networks, and new visualizations have been developed as a result, \eg~\cite{dedominico2015MuxViz}.

\paragraph*{Data Definition.} This aspect of the review looks at the nomenclature used for the dataset \eg multiplex, heterogeneous, which aspects are used to define layers across the data, as well as the structure of the data.

\paragraph*{Visualization Approach.} We analyse and categorise the various visualization approaches described, identifying novel approaches and novel applications of existing approaches \eg~\cite{bourqui2016multilayer}.
While many visualization systems described in this survey were not explicitly identified in the original source as being for multilayer networks, we point out ways in which they may be applicable and targeted to them.

\paragraph*{Interaction Approach:} Interaction with multiple layers will often be more complex and requires innovative techniques, such as~\cite{Hascoet:2012:IGM:2254556.2254654, shi2014Hierarchical,renoust2015detangler}.

\paragraph*{Attribute visualization:} Multilayer networks can also carry multivariate data~\cite{schreiber2014heterogeneous,dunne2012graphtrail}. Under this category we will examine the impact of multilayer structure on attribute visualization.

\paragraph*{Empirical Evaluation:} Empirical evaluation is a challenge for information visualization~\cite{Plaisant2004Evaluation}. Within the domain there are many guides to evaluation such as~\cite{purchase2012experimental}. However, techniques developed in application domains may not have been exposed to the same level of rigour as those developed within the visualization domain. It is important to understand which novel techniques have been empirically validated with respect to their usability.

\subsection{Papers Selection}
The wide range of application domains makes performing a complete survey highly challenging.
Within the domain of visualization, we queried prominent journals and conferences for a list of keywords related to multilayer graphs. Our main search engines were \textit{IEEE Explore} and the \textit{ACM Digital Library}.
The list included the terms (and variants of the terms using hyphens) multilayer, 
multilevel, faceted, multirelational, multimodal, multiplex, heterogeneous, and multidimensional.
The ambiguity of some of these terms meant that some completely unrelated papers were returned.
These were removed from the list based on their abstract.
The prominent visualization venues included \textit{IEEE TVCG} (and implicitly \textit{VAST} and \textit{Infovis}), \textit{CHI} (including \textit{SIGCHI} and \textit{TOCHI}), \textit{Computer Graphics Forum} (and implicitly \textit{Eurovis}), \textit{Advanced Visual Interfaces}, PacificVis, \textit{Graph Drawing and Network Visualization} (formerly \textit{Graph Drawing}), and the journal \textit{Information Visualization}.

Due to the wide range of application domains and numerous publication venues in each, it was not feasible to perform such a formalised search within them.
We used our initial list of visualization papers, as a seed adding papers form the application domains which were cited by or cited them as found using Google scholar search.

\FMrevised{Additional papers were also added to the list of those reviewed based on feedback from reviewers of this STAR, if they indicated that the papers would be valuable additions. Each paper was reviewed by at least 1 author, and the review shared with all other authors using a wiki. Papers were summarised based on the characteristics described in Section \ref{sect:categorisation}. Reviews of the paper were discussed at group meetings between the co-authors to provide a final decision on which papers should be included or excluded. All final text describing the papers within this work was validated by all co-authors.}

As stated in Section~\ref{sect:intro}, the goal of this survey is to reconcile the many visualization
approaches from the information visualization field and the application
domains. Many techniques have been extracted from papers which may not have focused explicitly on multilayer techniques, perhaps using one of the the names described in Section~\ref{sect:intro}, \eg heterogeneous.
However, the techniques are included as we believe that they are of interest to researchers who wish to visualize multilayer networks. 
\FMrevised{As part of the review process some papers were considered, based upon the keyword search described above, however, they were omitted from the final state of the art report due to their content not being related enough to the visualization of multilayer networks.}

\section{Survey of Multilayer Graph Visualizations}
\label{sect:thesurvey}
In this section we define and illustrate a task taxonomy for multilayer graphs.
Consistently with Munzner's model, we survey various data definitions on which the visualizations presented hereafter are built, as well as relevant interaction techniques. The survey encompasses the visualization of attributes in the context of multilayer networks and closes with considerations about visualization evaluation.
\subsection{Tasks And Analysis}
\label{subsec:task_analysis}
Numerous literature surveys \cite{lee2006task,Ahn2013Task,pretorius2014tasks, Kerracher:2015ik,beck2014state} list tasks relevant to the visual analysis of different types of networks (general, evolving, multivariate, \emph{etc}.) and tasks have been proposed on a domain specific basis, \eg~\cite{Murray2017Taxonomy}.

Lee \emph{et al.}~\cite{lee2006task} provide a general graph task taxonomy.
 At its top level it considers \textit{Topology Based Tasks}, \textit{Attribute Based Tasks}, \textit{Browsing Tasks}, and \textit{Overview Tasks}.
It explicitly specifies that the high level tasks  of comparison of graphs and identifying graph change over time are not covered by the taxonomy. 

Pretorius \emph{et al.}~\cite{pretorius2014tasks} focuses on multivariate networks.
The highest level of their taxonomy divides tasks as follows: \textit{Structure Based Tasks}, \textit{Attribute Based Tasks}, \textit{Browsing Tasks}, and \textit{Estimation Tasks}.
The category \textit{Estimation Tasks} is further subdivided and more detailed than Lee \emph{et al.}'s \textit{Overview Tasks}.
The name was chosen to capture that these tasks are not easily definable using lower level tasks and are considered more high level, and are not focused on giving precise answers.
Within this categorisation there is a comparison task, which may be of some relevance for multilayer graphs. It covers comparing information at different stages of a networks development, and determining causation, \emph{i.e.}, providing an explanation for the differences between  two snapshots of a changing network.

While Pretorius \emph{et al.} do consider graph change as part of their multivariate tasks taxonomy, the taxonomies of Kerracher \emph{et al.}~\cite{Kerracher:2015ik} and Ahn \emph{et al.}~\cite{Ahn2013Task} both focus  specifically on dynamic networks, also known as evolving or temporal networks.
At the highest level Ahn \emph{et al.}'s taxonomy focuses on three groupings: \textit{Entities}, \textit{Properties} and \textit{Temporal Features}.
The temporal features are grouped as \textit{Individual Events}, the \textit{Shape of Change} and the \textit{Rate of Change}.
These are considered from the individual entity level to the entire network level, and for both structural and domain properties.
Kerracher \emph{et al.}'s taxonomy builds on the non-network specific taxonomy of Adrienko and Adrienko~\cite{andrienko2006exploratory} by extending it to include network data.
It considers both elementary and synoptic tasks, as defined by Andrienko and Andrienko (elementary tasks involve individual items and characteristics, synoptic involve sets of items considered as an entity), but further divides synoptic tasks into three categories.
These are tasks considering graph subsets, tasks considering temporal subsets, and tasks considering  both graph and temporal subsets.
The taxonomy differs from Ahn \emph{et al.}'s in that it focuses more on the tasks that data items take part in, rather than the data items themselves, and considers a more general concept of pattern changes that captures relational changes in the network, as well as considering tasks which provide context for graph evolution.

Murray \emph{et al.}~\cite{Murray2017Taxonomy} propose a taxonomy in the context of biological pathway visualization that contains tasks concerning comparison, attribute analysis, and annotation that relate to multilayer networks.
Although most task taxonomies that have been developed so far do not directly address multilayer networks \emph{per se}, they could be further adapted or extended to target multilayer network visualization. 
Existing literature does mention specific tasks that may be relevant for multilayer network visualizations, which we cover in this section.
Some tasks may involve the temporal dimension as well (such as tracking the evolution of nodes or edges at different moments).

Unsurprisingly, tasks that are specific to multilayer networks revolve around the notion of a layer.
Tasks often boil down to manipulating elements within one layer, or across several layers, or manipulate the layers themselves.
These manipulations often lead to lower level tasks, which are also critical for visual analytics tasks (identifying actor roles, grasping group interaction or communication patterns in social networks, \emph{etc}.).

In the survey work of Pretorius \emph{et al.}~\cite{pretorius2014tasks}, a task is schematised as a process:

\emph{Select entity} $\to$ \emph{Select property} $\to$ \emph{Perform analytic activity}

We see here an important difference with the process of performing a task on a multilayer network involving layers.
Conceptually speaking, layers are genuine building blocks of a multilayer network.
They are neither a simple (sub-)network nor a mere property of a node or edge.
They are a conceptual construct that fully enters the analytical process when performing a task (involving the multilayer nature of the network).


We report here on different approaches or systems that support tasks relevant to multilayer networks. In many cases, authors have not explicitly expressed tasks in terms of layers, but rather referring to properties of the data they consider. This is the case for authors considering tasks related to group comparison or reconfiguration~\cite{Hascoet:2012:IGM:2254556.2254654,Cao:2015dy}. To this end, in anticipation of Section~\ref{subsec:discussion_taxonomy}, we propose task categories specific to multilayer networks. We target tasks directly involving visualization, as opposed to tasks that can be addressed through computational means only.

\begin{wrapfigure}{L}{0.4\columnwidth}
\includegraphics[width=0.4\columnwidth,keepaspectratio]{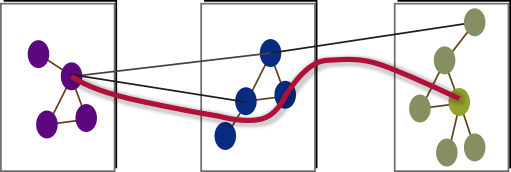}
\end{wrapfigure}
\noindent Task category~\textbf{A} - \emph{Cross layer entity connectivity} (\eg inter-layer path). Tasks in this category aim at exploring and/or inspecting connectivity involving paths traversing multiple layers. Understanding how shortest paths expand across layers, inspecting what nodes do occur on these paths are typical examples of tasks in this category. Being able to explore cross layer connectivity has been identified as an important user task in~\cite{Ghani:2013fo}. Associative browsing in Refinery~\cite{kairam2015refinery} is a good illustration of cross layer connectivity task. It performs cross-layer random walks and collects nodes from different layers in a single view. The leapfrogging operation in Detangler~\cite{renoust2015detangler} is another good illustration of cross layer connectivity building a dual view reflecting how/what layers get involved when hopping from node to node (see Section~\ref{subsec:interaction}).

\begin{wrapfigure}{L}{0.4\columnwidth}
\includegraphics[width=0.4\columnwidth,keepaspectratio]{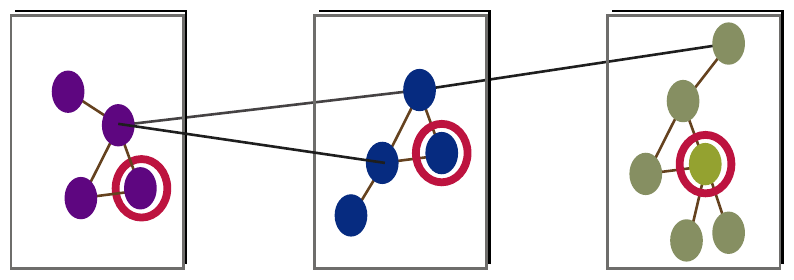}
\end{wrapfigure}
\noindent Task category~\textbf{B} - \emph{Cross layer entity comparison}. Tasks in this category aim at comparing entities (typically, nodes) across different layers; this requires the ability to query entities across layers. The task may concern the same (set of) node(s) over several layers; or distinct nodes that are somehow linked across different layers. Jigsaw~\cite{stasko2008Jigsaw} typically supports this tasks by allowing users to identify entities (persons, places, etc.) through several documents (seen as layers in a multilayer document network). \BPrevised{FacetAtlas~\cite{cao2010facetatlas}} multi-facet query box is another good example.
\vspace{8pt}

\begin{wrapfigure}{L}{0.4\columnwidth}
\includegraphics[width=0.4\columnwidth,keepaspectratio]{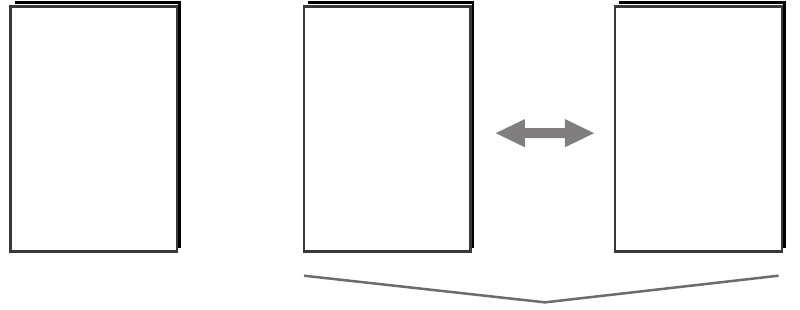}
\end{wrapfigure}
\noindent Task category~\textbf{C} - \emph{Layer manipulation, reconfiguration (split, merge, clone, project)}. Tasks in this category aim at manipulating the layer structure itself. Such manipulation may allow for previously unseen relationships and structure to be revealed, and allow for new perspectives on the underlying data. Combining layers through drag \& drop operations as in \cite{Hascoet:2012:IGM:2254556.2254654} is a perfect illustration of this type of tasks; another example is g-Miner~\cite{Cao:2015dy} which allows to create, edit or refine the grouping of elements.
\vspace{8pt}

\begin{wrapfigure}{L}{0.4\columnwidth}
\includegraphics[width=0.4\columnwidth,keepaspectratio]{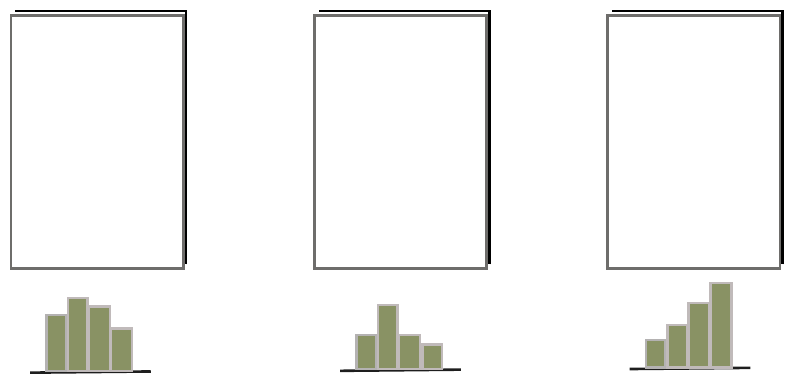}
\end{wrapfigure}
\noindent Task category~\textbf{D1} - \emph{Layer comparison based on numerical  attributes}. Tasks in this category support comparing layers to one another based on numerical measures summarising layer content and structure. Typically, layers could be compared by looking at how node degree distributions compare layer-wise. OntoVis~\cite{shen2006} (where layers map to node type) support layer comparison tasks using a metric they call (inter-layer) \emph{node disparity}. Pretorius \emph{et al.}~\cite{pretorius2008} propose a quite elaborate approach and system to perform multi-attribute-based layer comparison.
\vspace{8pt}

\begin{wrapfigure}{L}{0.4\columnwidth}
\includegraphics[width=0.4\columnwidth,keepaspectratio]{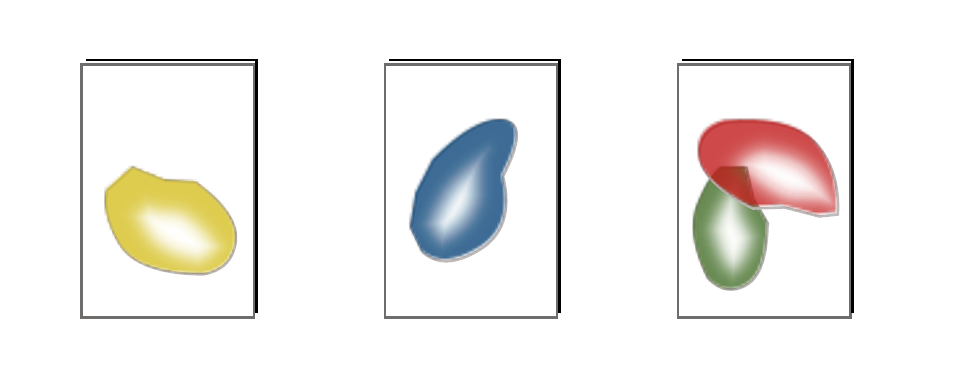}
\end{wrapfigure}
\noindent Task category~\textbf{D2} - \emph{Layer comparison based on topological, connectivity patterns, layer interaction}. Tasks in this category support comparing layers through non-numerical but rather topological features of layers (\eg group structure). A layer could be  hierarchical (inheritance), while another could show a strong scale-free structure, for instance. The work by Vehlow \emph{et al.}~\cite{Vehlow:2015gk} is a typical technique allowing to compare group structure across layers. Tasks R5 and R12 in GraphDice~\cite{Bezerianos:2010vb} are another good illustration of such tasks.

Table \ref{table:tasks} summarises task categories supported by a selection of systems and techniques cited and described in this report.

\begin{table*}[!ht]
 \centering
 \begin{tabular}{r|c|c|c|c|c|}
 \cline{2-6}
 
 &\textbf{A} - Cross layer&\textbf{B} - Cross layer&\textbf{C} - Layer manip.&\multicolumn{2}{c|}{\textbf{D} - Layer comparison}\\
  & \shortstack[c]{connectivity}&  \shortstack[c]{entity comparison} &  \shortstack[c]{reconfiguration} & \shortstack[c]{\textbf{D1}\\ numerical}&\shortstack[c]{\textbf{D2}\\ topological}\\
 \cline{2-6}
GraphDice \cite{Bezerianos:2010vb} & & \shortstack[c]{\OK \\ {\footnotesize (multi facet query)}} & & & \BPrevised{\shortstack[c]{\OK \\ {\footnotesize (R5, R12)}}}\\
\cmidrule{2-6}
\shortstack[c]{Multilayer Graph\\ Edge Bundling~\cite{bourqui2016multilayer}}& \OK & \OK &&& \\
\cmidrule{2-6}
VisLink~\cite{collins2007vislink} & \OK &&&& \\
\cmidrule{2-6}
\BPrevised{g-Miner} \cite{Cao:2015dy} & &  & \OK & \multicolumn{2}{c|}{\OK}\\
\cmidrule{2-6}
FacetAtlas \cite{cao2010facetatlas} && \BPrevised{\OK} &&& \OK \\
\cmidrule{2-6}
MuxViz \cite{dedominico2015MuxViz} & \OK & \OK & &  \multicolumn{2}{c|}{\OK}\\
\cmidrule{2-6}
GraphTrail \cite{dunne2012graphtrail} & & \OK &&&\\
\cmidrule{2-6}
ManyNets \cite{freire2010ManyNets} & & \OK & & &\\
\cmidrule{2-6}
\shortstack[c]{Multimododal\\ Social Networks \cite{Ghani:2013fo}} & \shortstack[c]{\OK \\ (Q1b,c)} & \shortstack[c]{\OK \\ (Q1a)} & \OK & & \shortstack[c]{\OK \\ (Q2c)} \\
\cmidrule{2-6}
Donatien \cite{Hascoet:2012:IGM:2254556.2254654} & & & \OK && \OK \\
\cmidrule{2-6}
\shortstack[c]{Hierarchical\\ Edge Bundling~\cite{holten2008}} & & \OK & & & \OK \\
\cmidrule{2-6}
Hive Plots~\cite{krzywinski2011hive}& & & \OK & \OK & \\
\cmidrule{2-6}
Refinery \cite{kairam2015refinery} & \shortstack[c]{\OK \\ {\footnotesize (assoc. browsing)}} & & & & \OK \\
\cmidrule{2-6}
Circos~\cite{krzywinski2009circos}& & & & \OK &\\
\cmidrule{2-6}
HybridVis~\cite{Liu:2017it}& \shortstack[c]{\OK \\ (Q4)} & & \shortstack[c]{\OK \\ (Q1, Q2)} & \multicolumn{2}{c|}{\shortstack[c]{\OK \\ (Q3)}} \\
\cmidrule{2-6}
Detangler \cite{renoust2015detangler} & \shortstack[c]{\OK \\{\footnotesize (leap-frogging)}} & & \OK & \multicolumn{2}{c|}{\OK} \\
\cmidrule{2-6}
Jigsaw \cite{stasko2008Jigsaw} &&&& \multicolumn{2}{c|}{\shortstack[c]{\OK \\ (disparity)}}  \\
\cmidrule{2-6}
Ontovis~\cite{shen2006}& & & \OK &  \multicolumn{2}{c|}{\OK} \\
\cmidrule{2-6}
BicOverlapper~\cite{santamaria2008visual}& \OK & \OK & & & \OK \\
\cmidrule{2-6}
\shortstack[c]{Dynamic\\ communities \cite{Vehlow:2015gk} }&&& \OK && \OK\\
\cmidrule{2-6}
Pivot Graphs \cite{Wattenberg:2006js} &  & & & &\shortstack[c]{\OK \\ (roll-up)}\\
\cmidrule{2-6}
NetworkAnalyst tool~\cite{xia2015networkanalyst} & & \OK & \OK & &\\
\cline{2-6}
 \end{tabular}
  \caption{A selection of techniques/systems (bibliographic order) mapped onto tasks categories, relevant to multilayer networks, that they either implicitly or explicitly support. Notes in parentheses refer to task labelling/naming as indicated by authors in their  paper.}
	\label{table:tasks}
\end{table*}

\subsection{Data Definition}
\label{subsec:data_definition}
This subsection looks at the various data definitions found in the visualization literature on which visual representations of networks with multilayer characteristics are built. Only a few approaches explicitly mention the use of multilayer networks (both as data underlying the visualization and as a visual encoding). Most systems dealing with multivariate networks couple relational data with node and edge attributes \cite{shen2006,Wattenberg:2006js,Bezerianos:2010vb,heer2014orion} often using table-based representations \cite{Kerren2014,heer2014orion}; they do not consider any data or attribute specifying a layer structure. 
Cao \emph{et al.}~\cite{cao2010facetatlas} consider \emph{classes} of entities they call ``\emph{facet}'' which appear naturally map to layers of nodes (see Section \ref{sec:multivariate}). Among all, the work of Pretorius \emph{et al.}~\cite{pretorius2008} is a notable exception as it introduces the notion of layers without using the term, and explicitly defines nodes as Cartesian products of attributes (see Section~\ref{subsec:layer_related_concepts}).

Other systems and approaches infer multilayer structure by aggregating data from multiple sources, whether databases \cite{kohlbacher2014multivariate} or a collection of ego networks (as in \cite{DUNBAR2015}) and/or personal data \cite{huang2015}. Interestingly enough, some systems do not directly target the visualization of multilayer networks, but use multiplex and/or hypergraph representations to build query graphs or summarise query response~\cite{tu2013,Shadoan2013}.

Obviously, MuxViz \cite{dedominico2015MuxViz} relies on the exact definition and implementation (see Section~\ref{subsec:layer_related_concepts}) introduced by~\cite{MultilayerNetworks}, which is also the case of authors mentioning explicit use of the MuxViz framework~\cite{Barthelemey15}. Elementary layers originating from aspects of the network, such as time or node/edge type, are quite similar to the facets described  in~\cite{hadlak2015survey}.
Detangler~\cite{renoust2015detangler} relies on an explicit encoding of layers,
with a goal to allow an easy exploration of inter-layer correlation (see Section~\ref{subsec:task_analysis}). Making a distinction between layers as being either structural or functional (or of any other type) may be useful depending on the pursued goal~\cite{Agarwal2017}.

\subsection{Visualization Approaches}
\label{subsec:survey_visualization}
From a multilayer network perspective, previous work in network visualization techniques may be classified based on their awareness of the notion of a layer. When this is the case, layers are visually encoded using any appropriate Gestalt principle in a way that structures the spatial representation; they are also manipulated as visual objects in their own right as detailed in Section~\ref{subsec:task_analysis}.
This is why this section is organised based on the type of visual encoding used to show layers explicitly. This survey also documents and reflects on the widespread use of weaker visual cues (in the sense of Mackinlay's ranking of perceptual tasks~\cite{mackinlay1986}) to encode layer information, such as node or link colour.

\subsubsection{1-Dimensional Representations of Layers} 
Existing visualization techniques use a large variety of one-dimensional representations of layers.
This type of visual encoding relies on the \emph{law of continuation} of Gestalt theory, such that the eye may perceive paths on which nodes are arranged whether theses paths are actually drawn or not.
This applies to circular paths, as well as straight axes, or any curve shape.

\paragraph*{Circular Representations.}
This body of work includes concentric circles, where each circle stands for a layer.
Concentric circles are used in~\cite{bothorel2013visualization} where the focus is on depicting paths through the whole set of layers (Task category~\textbf{A} in our taxonomy).
Node order optimisation and edge bundling are used to reduce edge clutter.
A similar layout is used in the ring view of MuxViz~\cite{dedominico2015MuxViz} but focuses on visual correlation analysis of node attributes across different layers (Task category~\textbf{D1}). Node colour encodes attribute values (see Section~\ref{subsec:attrib_vis} and Figure~\ref{fig:muxviz_rings}), while ring order and ring thickness encode computed layer-level metrics.
Similarly, Circos~\cite{krzywinski2009circos} is a popular tool for comparative analysis of genomic data, where each ring/layer may stand for a biological sample. In order to compare node attribute values across samples, a histogram is wrapped around each ring (Task category~\textbf{D1}). 

Chord diagrams display layers as arcs composing one overall circle. They are used in the NetworkAnalyst tool~\cite{xia2015networkanalyst} to analyse gene expression data. Links between layers are drawn as splines connecting identical nodes occurring in different layers/arcs (Task category~\textbf{B}). The analyst may click on a pair of arcs to highlight their common nodes (and the bridging links). A similar approach is followed in~\cite{alsallakh2013radial,crnovrsanin2014visualization}. In presence of multilevel categorical attributes as in~\cite{humayoun2016social}, each arc of the chord diagram can further be split hierarchically (Task category~\textbf{C}). The chords would then connect nodes at the leaf level across all layers where they are repeated.

\paragraph*{Axis-based Node-Link Representations.}
In this category a layer is materialised by a straight 1-dimensional axis. Obviously, the representation of a multilayer network lays out nodes on several such parallel axes. An important way of distinguishing axis-based visualizations relates to the type of variable represented by the axis, whether it is quantitative, \eg graph metric like node degree or any numeric node attribute, or ordinal/ranking-based.
Despite the visual similarity to the Parallel Coordinates plot~\cite{inselberg1990parallel}, a polyline represents a path between nodes sitting in different layers/axes, rather than a thread linking attribute values across different columns in a given table entry. 
Crnovrsanin \emph{et al.}~\cite{crnovrsanin2014visualization} describe a view that uses such parallel axes arrangement, and alternatively chord diagrams. 
An example of analyses they run consists in comparing the ``aggression network'' among students in four different schools, based on student race group.
They show that smaller groups do not show internal aggression patterns, while larger groups victimise everybody equally (within the same group and in other groups).
In this case the analyst is more interested by topological considerations at the group level, and structural differences between layers (Task category~\textbf{D2}).

\begin{figure}[!ht]
  \centering
   \includegraphics[width=\columnwidth]{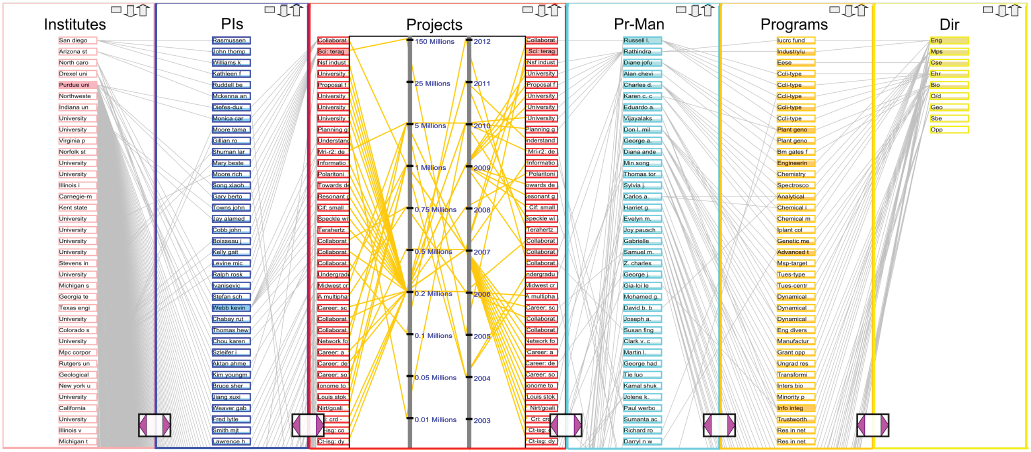}
  \caption{The PNLB (Parallel Node Link Bands) representation of~\cite{Ghani:2013fo}. Each axis is a distinct set of vertices. Edges are only displayed between adjacent axes. Some axes show a quantitative value \eg project budget, while others display text strings sorted based on a graph metric or alphabetically.}
  \label{fig:PNLbs}
\end{figure}

Ghani \emph{et al.}~\cite{Ghani:2013fo} provided an approach called Parallel Node Link Bands (PNLBs). Nodes are positioned uniformly across spaced parallel axes which represent layers defined by the node type (or mode), see Figure~\ref{fig:PNLbs}.
Edges are only drawn between adjacent layers, and within layer edges are shown in a separate visualization. 
Node order on axes can be set based on edge attributes or connectivity to other layers.
They use their approach to analyse the NSF funding dataset.
Examples of tasks they carry out include determining whether some NSF program manager award funding to some PIs more often than others on a 3-layer networking containing program managers, projects, and PIs.
This is an instance of Task category~\textbf{A} where the focus is on paths traversing all layers.

The list view of Jigsaw\cite{stasko2008Jigsaw} provides an overview of entities grouped by type, with edges being drawn between connected entities in adjacent lists.
One of the main utilities of this system is to relate different types of named entities (people, geographic locations, organisations) mentioned in the same documents.
Entities which are connected to a currently selected item are highlighted by colour across all lists.
It therefore emphasises the analysis of paths across all available layers (Task category~\textbf{A}).
The list view is complemented by a node-link and a matrix-like scatterplot view amongst others.

\begin{figure}[!ht]
  \centering
   \includegraphics[width=\columnwidth]{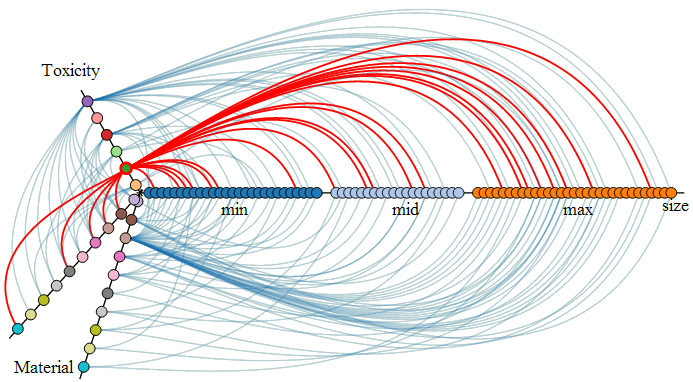}
  \caption{The hive plot representation of health data by~\cite{yang2016user} showing 4 layers/axes: toxicity type (duplicated), material and particle size. Edges are only displayed between adjacent axes. The vertices on the horizontal axis are coloured based on their cluster membership.}
  \label{fig:hive}
\end{figure}

The Hive Plots~\cite{krzywinski2011hive} differ from the previous techniques in that they arrange the axes radially. 
Originally introduced for the analysis of genomic data, they have been used in other domains like performance tuning in distributed computing~\cite{engle2012visualizing} and in the domain of health~\cite{yang2016user} as can be seen in Figure~\ref{fig:hive}. In~\cite{krzywinski2011hive}, node (gene) subsets are placed on separate axes based on a node partitioning algorithm.
The fundamental questions they answer using Hive Plots include determining differences in connectivity patterns between layers (Task category~\textbf{D1}).
An element's position along its axis is often calculated based on a graph metric, \eg node degree in~\cite{engle2012visualizing}and may be based on the raw or normalised value of an attribute.
Edges are displayed between adjacent axes only.
Yet, visual clutter may still occur with real application data.
Layer duplication as in Figure~\ref{fig:hive} is convenient when the relationship to a non adjacent axis becomes necessary (Task category~\textbf{C}).

\subsubsection{2D, 2.5D and 3D Node-Link Representations}
Across the various papers we surveyed, node-link layouts cropped up frequently.
The MuxViz toolkit\cite{dedominico2015MuxViz}, from the domain of complex systems, utilises standard node-link visualizations.
They are also used in other domains that depend on complex systems theory\cite{Barthelemey15,Bentley2016,deDomenico2017}. 

\begin{figure}[!ht]
	\centering
		\includegraphics[width=0.9\columnwidth]{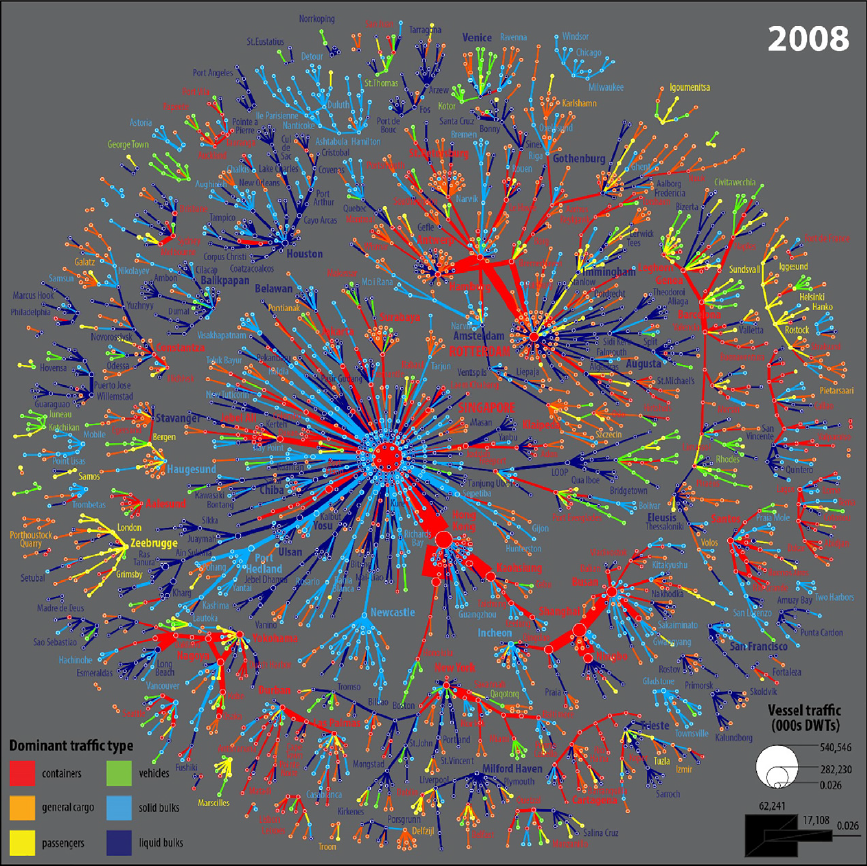}
	\caption{A multilayer network visualization describing the flow of maritime traffic. Nodes represent ports and different edge colours represent different modes of shipping, taken from~\cite{ducruet2017Maritime}.}
	\label{fig:maritime_flows_ducruet}
\end{figure}

A widespread visual design consists in encoding layer information using node colour or shape, as depicted in~\cite{moody2005dynamic,fung2009visual,kohlbacher2014multivariate,Zeng:2016ez}.
Colour coding of edges is also used in~\cite{ducruet2017Maritime,dedominico2015MuxViz}.
This design choice relies on the law of similarity of Gestalt theory (colour similarity in this case). 
This design is often adopted when the multivariate nature of the network is the driving motivation of the visual design. 
For instance, Figure~\ref{fig:maritime_flows_ducruet} represents flows of maritime traffic using colour to encode different modes of shipping (or layers). 
The analyst looks among other things at structural changes over time, where different layers encode different time slices (Task category~\textbf{D2}).
But if the analyst is interested in analysing a given time slice, different layers may represent different shipping modes.
The related task consists in comparing structural differences among the different modes.
In similar visual designs, layer information is diffuse, relationships between layers and within the same layer are mixed and users seldom get a handle on layers to manipulate them directly.
Nodes belonging to different layers are intertwined in the 2D plane, when standard node-link layouts are used, and edge clutter is problematic.
Layer-related tasks may therefore be difficult to carry out under these circumstances.

While not explicitly designed with multilayer network visualization in mind, constraint based layouts offer the possibility to constrain a two dimensional node-link layout in such a way that respects the concept of layers. For example the \emph{SetCola} constraint-based layout of Hoffswell \emph{et al.}~\cite{hoffswell2018setcola} allows users to apply layout constraints to sets of nodes, which might easily correspond to layers. Such a layout approach supports analysing cross layer connectivity (Task category~\textbf{A}) as well as layer comparison (Task category~\textbf{D2}). The examples covered by the authors include a food web networks and a network modelling a biological cell, and both of these datasets can be considered to have multilayer characteristics.

Inspired by the multi-level nature of some problem areas \eg biological networks, the 2.5D approach materialises layers as 2D translucent parallel planes in a three dimensional layout, similar in spirit to Figure~\ref{fig:layer_illustration}.
This visual design relies on the law of uniform connectedness of Gestalt theory.
It separates links lying within layer from those between layers providing a more natural support for path related tasks (Task categories~\textbf{A} and~\textbf{B}) than traditional 2D node-link layouts, but 3D navigation is required to allow the user to change his perspective on the data and resolve visual occlusion problems.
As opposed to 1D axis-based representations, the parallel 2D planes provide space to lay out intra-layer links.
In the 2.5D category, some approaches use colour redundantly to encode layer information as in~\cite{fung2009visual}.
Other visual design options for 2.5D consist in using colour to encode an attribute value or a computed metric, \eg community assignment by a community detection algorithm as in~\cite{dedominico2015MuxViz}, across the different layers.
From the biological domain, the \emph{Arena3D} application visualizes biological data using an interactive 3D layout, where layers are also projected onto planes, and entities are connected across layers by edges rendered as 3D tubes. The authors demonstrate its effectiveness by analysing  the relationship across layers, based on proteins and genes associated with a specific disease (Task category~\textbf{A}).

The use of three dimensional layouts is much less common in the information visualization research community. While some work has shown that there may be some benefit to three dimensional layouts, this is only under stereoscopic viewing conditions~\cite{Ware2008}.
Outside of stereoscopic viewing conditions, there are no empirical studies which demonstrate usability gains from a three dimensional graph visualization~\cite{greffard2011visual}.

A more widely accepted approach in information visualization, especially for the purpose of comparative analysis of graphs, consists in using small multiple views. This is often used for graph matching tasks, where the focus is on understanding commonalities and differences between a set of related networks~\cite{Hascoet:2012:IGM:2254556.2254654}. In the context of this paper, the networks that need to be matched are distinct layers in a larger multilayer network (Task categories~\textbf{D1} and~\textbf{D2}). Whether in a 2.5D setting or in a flat small multiples setting, one challenge consists in ensuring that duplicate nodes are laid out consistently across layers, by introducing constrained layout strategies as in~\cite{fung2009visual,Hascoet:2012:IGM:2254556.2254654} to better support cross layer entity comparison (Task category~\textbf{B}).

More generally, coordinated multiple views are often used in the domain of information visualization, and in many applications \eg the analysis of microarray data~\cite{santamaria2008visual}. In this case, two-dimensional node-link views may be used as one of multiple complementary visualizations of a multilayer network, \eg~\cite{kairam2015refinery,Ghani:2013fo,stasko2008Jigsaw}. It is yet possible to eschew the idea of using a node-link visualization altogether~\cite{dunne2012graphtrail}.
Coordination between views is common, \eg brushing and linking. The Detangler~\cite{renoust2015detangler} application builds on this by also harmonising layouts between views.
It supports several task categories identified in this survey, namely cross layer connectivity (Task category~\textbf{A}), layer manipulation (Task category~\textbf{C}) and layer comparison (Task categories~\textbf{D1} and~\textbf{D2}).

\begin{figure}[!ht]
  \centering
   \includegraphics[width=\columnwidth]{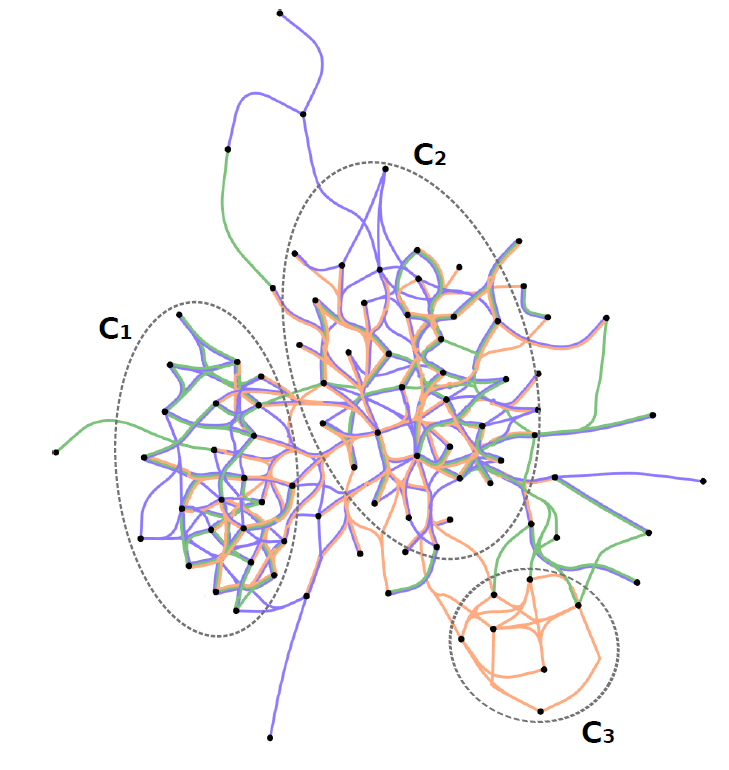}
  \caption{The multilayer edge bundling of~\cite{bourqui2016multilayer}}
  \label{fig:multilayer_edge_bundling}
\end{figure}

\paragraph*{Edge Visualization}
The complex structure of multilayer graphs makes edge visualization an important challenge.
It may be important in some cases to distinguish between inter-layer and intra-layer links, in other cases the number of layers may cause enough clutter with respect to edges, that a visualization becomes less understandable.
In some cases, the chosen solutions is to simply not draw all edges and to allow the user to choose which edges to see via interaction to ease inter-layer comparisons (Task category \textbf{B}). For example, the PNLB (Parallel Node Link Bands) technique~\cite{Ghani:2013fo} only draws inter-layer edges between nodes on parallel axes, and intra-layer edges are displayed in a separate visualization.
The well established technique of edge bundling~\cite{holten2006} has been adapted for the multilayer use case by~\cite{bourqui2016multilayer}. 
The authors bundle all edges in a single visualization, in an aesthetically pleasing manner, with edges being kept adjacent to each other when they share a common path, and edge crossing being avoided (see Figure~\ref{fig:multilayer_edge_bundling}). This approach is useful for showing edges from multiple layers in a single visualization (where there is no division of nodes between layers); the approach is agnostic to the source or target layer, or whether the edges are between or within layer(Task categories \textbf{A} and \textbf{B}).

Within their list-based view \cite{crnovrsanin2014visualization} use edge bundling between different list columns as a clutter reduction techniques clarify similarities between different edge types.
The authors essentially group edges based on relation type, by clustering the vertices and altering the clustering based on vertex mode.
\begin{figure}[!ht]
  \centering
   \includegraphics[width=0.8\columnwidth]{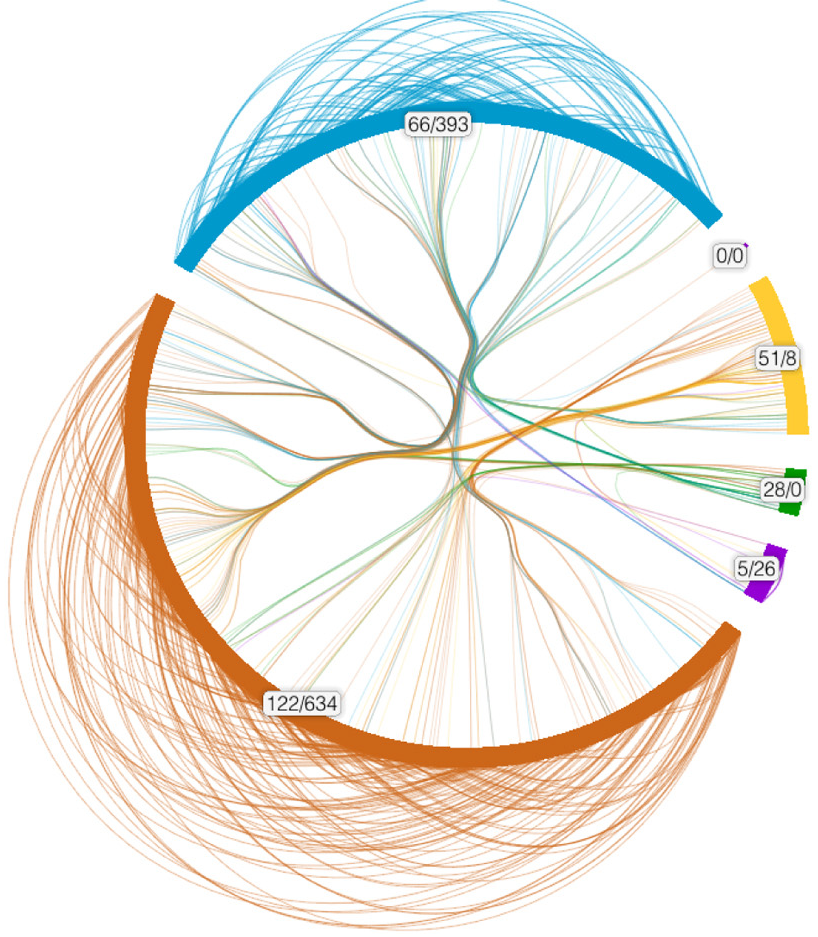}
  \caption{Edge bundling as utilised by~\cite{crnovrsanin2014visualization}. Within category edges are routed around the exterior of the circle. Between category  edges are routed via the interior of the circle and bundled.}
  \label{fig:multi-edge}
\end{figure}
They also use a modified edge bundling in their  circular layout,  that distinguishes within-mode edges and between-mode edges, see Figure~\ref{fig:multi-edge}. 

\subsubsection{Matrix-Based Visualizations}
Standard node-link representations of graphs give equal importance to nodes and links and aim usually to convey structural properties of the graph at hand.
They may however be difficult to read due to edge clutter for moderate size graphs, and for more complex networks encountered in many real usage scenarios.
When dealing with large and/or dense graphs, matrix-based representations were found to be more readable than node-link diagrams~\cite{ghoniem2005readability} for many tasks, except path finding.
They consist in laying out nodes as the rows and columns of a 2-way table.
A link between two nodes is often represented as a rectangle at the intersection of the associated row and column.
This avoids altogether the edge clutter problem of standard node-link representations.
Colour is often used to encode the weight of the links, when link attribute values are available.
This makes matrices very similar, if not identical in essence, to heatmap views frequently used in biology and other domains~\cite{wilkinson2009history}.
Other visual designs include using circles at the intersection of rows and columns with size and colour encoding link attribute values, as in~\cite{Chuang:2012:TVT:2254556.2254572}.
Matrix representations have been used to visualize homogeneous graphs (nodes of one type), \eg in software engineering~\cite{van2003using}, and bipartite (or 2-mode) graphs, \eg in software performance tuning~\cite{ghoniem2005peeking}.

The ability to detect link patterns in a matrix view is conditioned by the use of an appropriate ordering of rows and columns. 
Various seriation algorithms~\cite{chen2002generalized,liiv2010seriation,fekete2015reorder} reorder the rows and columns of the matrix to create dense rectangular blocks of links. 
Community detection in a bipartite graph consists in finding groups of nodes in one layer which are densely connected to groups of nodes found in the second layer (Task category~\textbf{A}).
2-way hierarchical clustering is commonly used with biological data for this purpose.
The BicOverlapper system~\cite{santamaria2008visual} uses biclustering methods to find such relationships between groups of genes and related groups of medical conditions.
On the visual side, BicOverlapper uses coordinated multiple views, one of which employs convex hulls within a standard node-link representation to materialise groups of genes, akin to the notion of elementary layers described in Section~\ref{subsec:data_definition}. The overlapping convex hulls are meant to support the identification of commonalities and differences between  layers (Task categories~\textbf{B} and~\textbf{D2}).

In presence of multiple layers, the comparison of link patterns between many pairs of layers may be useful to the analyst (Task category~\textbf{A}, see Section~\ref{subsec:task_analysis}). 
Laying out small multiples of matrix views side by side is one approach. Liu and Shen~\cite{Liu:2015bf} investigates several possible juxtaposition strategies, and assess their usability with multifaceted, time-varying networks.
MuxViz~\cite{dedominico2015MuxViz} uses matrices to summarise layer-level statistics, as a means to convey a notion of layer similarity to the analyst (Task categories~\textbf{A},~\textbf{B},~\textbf{D1}, and~\textbf{D2}).

\subsubsection{Hybrid Approaches}
\label{subsec:survey_visualization_hybrid}
Recent work has been exploring the integration of multiple visualization techniques, as an effort to better grasp underlying data~\cite{Javed:2012vt}.
Although matrices have been shown superior to node-link diagrams for dense networks, the latter may facilitate the tracking of edge directions. In this spirit, NodeTrix~\cite{Henry:2007er} mixes node-link views with matrix-based visualizations to support typically locally dense social networks.
While NodeTrix is not explicitly a multilayer network visualization technique, it is the first hybrid approach that focused specifically on network visualization. Since its inception, the idea has been extended by other techniques to support other types of data, such as compound graphs~\cite{rufiange2012treematrix}.
Although they do not always focus on visualizing multilayer networks, such approaches could also be directly reused or adapted to support multilayer networks.
VisLink~\cite{collins2007vislink}, for instance, allows visualizing a data set using multiple representations at once, also explicitly displaying the cross-views links.
Using the technique, one layer could be used for each representation, and inter-layer links could be highlighted (Task category~\textbf{A}).
Adopting another perspective, HybridVis~\cite{Liu:2017it} allows using the same kind of representation, but for different levels of details (or hierarchical scales). In this case, a node-link view may include some levels that are shown as expanded, and other levels are shown as collapsed (Task category~\textbf{C}). With additional views (histograms, parallel coordinates) more details on level attributes can also be obtained (Task categories~\textbf{D1} and~\textbf{D2}).

\subsection{Interaction Approaches}
\label{subsec:interaction}
The discussion about user interactions may be grounded in Yi \emph{et al.}'s categorisation of interaction techniques~\cite{yi2007toward}.
According to Hasco\"et and Dragicevic~\cite{Hascoet:2012:IGM:2254556.2254654}, multilayer network visualizations may support user interaction at the level of individual network elements (\eg individual nodes and links), and at the level of whole layers whether single layers or groups of layers (Task category~\textbf{C}). They argue that layer level interactions require a visual affordance. In particular their system, called Donatien, supports the Yi \emph{et al.} reconfigure and explore interactions. 

Traditional interactions include:
\begin{itemize}
	\item selection: point and click selection, lasso selection of nodes;
    \item filtering: keeping/removing nodes or links based on attribute values;
    \item navigation: to visually inspect a fragment of the visual representation using zoom and pan, or context+detail techniques (\eg fisheye distortion or magic lenses).   
\end{itemize}

These have obviously been used widely with standard node-link representations, and are directly applicable one layer at a time in the context of multilayer networks.
Interacting with entire layers is however more relevant to the present discussion and ties back to layer level tasks described earlier in Section~\ref{subsec:task_analysis} (Task categories~\textbf{D1} and~\textbf{D2}).
Donatien offers three different spatial organisations of layers:
\begin{enumerate*}
\item small multiples;
\item stacking the layers on top of each other; 
\item animation.
\end{enumerate*}
Starting from the small multiples view, the analyst can drag and drop a layer onto another one, to stack them and more easily compare their elements based on the distinctive layer colour. 
In the stacked mode, a set of title bars provides an affordance to reorder the layers in the stack interactively.
The title bars also include reconfiguration tools \eg choices of layout algorithms that are applied to the layer being manipulated or to the whole stack of layers.
Crossing-based interaction across the set of title bars is used to achieve flipbook animation, also for the sake of comparison across layers.
This seems quite a natural approach when the layers are defined as consecutive snapshots of a dynamic network.
Also in the stacked mode, Donatien clusters nodes from different layers together based on their spatial proximity in the pixel space.
The analyst is yet allowed to edit the resulting clusters interactively by pulling a node out, or by dragging and dropping a node on another node (or group of nodes) to merge them.
Merged nodes carry a colour coded pictogram relating them to the layers they occur in. 

More structured layer organisations may prove to be necessary \eg a hierarchy of layers.
This ensues from the concepts of aspects, layers and elementary layers put forth by Kivel\"a et al, but also to many real application needs.
From an interaction perspective, merging layers together or splitting them apart becomes a matter of collapsing or expanding their parent node in that layer hierarchy. 
In this vicinity, the Ontovis system~\cite{shen2006} uses an ontology visualization to steer the associated network visualization. An ontology could be seen as an artifact representing the layer structure of a multilayer network.

\begin{figure*}[!ht]
	\centering
		\includegraphics[width=\linewidth]{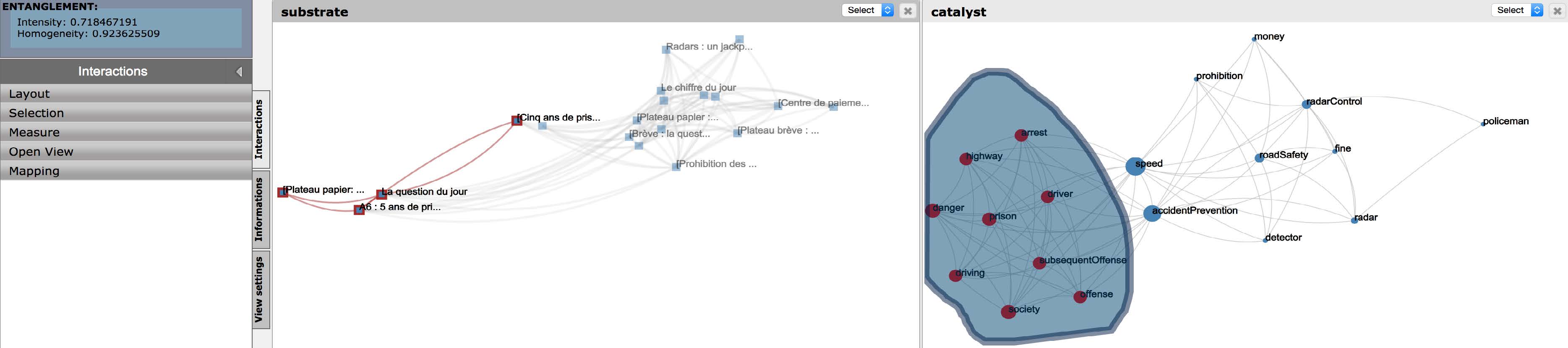}
	\caption{A screen shot from Detangler~\cite{renoust2015detangler} showing how nodes (left panel) relate to layers (right panel). Selecting layers (lasso) trigger the selection of nodes they involve (red nodes, left panel).}
	
	\label{fig:detangler}
\end{figure*}

The Detangler approach~\cite{renoust2015detangler} combines two distinct, synchronised visual representations. A first panel (Figure~\ref{fig:detangler}, left) displays the overall network connectivity through a node-link view between nodes of all layers. Another panel (right) displays a node-link view showing how layers interact (where interaction is measured and inferred in an \emph{ad hoc}, domain dependent, manner). Detangler supports a ``leapfrogging'' interaction: the selection of nodes in the left panel automatically triggers the corresponding layers in the right panel (Task category~\textbf{A}).  Leapfrogging (executed by double-clicking the selection lasso) \emph{expands} the original selection to include all nodes (Boolean OR) involved in any one of the layer that got selected; or \emph{restricts} the original selection to nodes involved in all layers that got selected from the layer view (Boolean AND).

The \emph{OnionGraph} application~\cite{shi2014Hierarchical} provides a hierarchical focus and context approach targeted specifically toward heterogeneous data.
The hierarchy provides different levels of abstraction based on node type, role equivalence and structural equivalence. 
In their example use case, using an academic publication dataset, the heterogeneity of the data is derived from node types, and edges only exist between certain node types. There is no formal layer definition and the abstraction used to provide the hierarchical focus and context abstraction is applied across all data types and does not fully consider the heterogeneity of the data.
Such abstractions could be adapted to be applied on a per layer basis. This could be very useful in multilayer systems, particularly for comparison of complex layers.

\subsection{Attribute visualization}
\label{subsec:attrib_vis}
\begin{figure}[!ht]
	\centering
		\includegraphics[width=0.8\columnwidth]{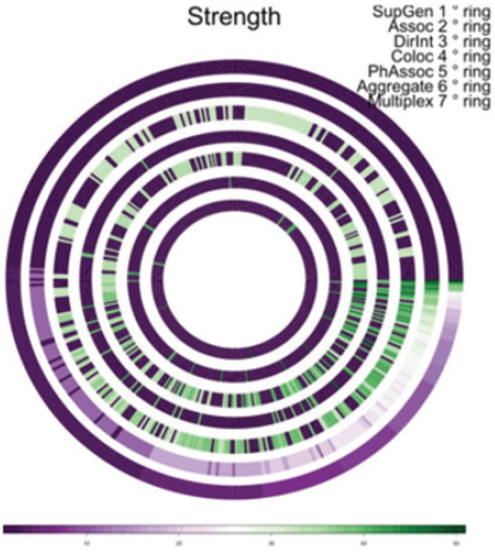}
	\caption{A screen shot from MuxViz~\cite{dedominico2015MuxViz} showing the values for a centrality across layers. Each ring specifies a different layer.}
	\label{fig:muxviz_rings}
\end{figure}

As with standard network data, node and edge entities in multilayer networks may have many attributes, either categorical of numerical, associated with them. 
However within a multilayer network, attributes of nodes are not only considered within a single network context. 
Attributes need to be considered across layers, and attribute values (especially for numerical attributes) may change across layers, especially if the attributes are derived from graph metrics, which may be calculated on a per layer basis.
An example of this can be seen in the MuxViz toolkit~\cite{dedominico2015MuxViz}.
Here the authors use an annular ring visualization approach, which show the values of metrics across layers, with each ring representing a layer, or in some cases a different centrality for a specific layer see Figure~\ref{fig:muxviz_rings} (Task category~\textbf{D1} or~\textbf{D2}). 

This basic approach involves completely separating the attribute visualization from the graph structure. To better relate the relationship between networks structure and attributes, the attributes may be integrated into the network visualization itself, (referred to as augmented network visualization by~\cite{Dickison2016,rossi2015}) or a linked view brushing approach maybe taken, by which the relevant related nodes would be highlighted in a network view, when selected in the attribute visualization and \emph{vice versa} (Task category~\textbf{B}).

\begin{figure}[!ht]
	\centering
	\includegraphics[width=\columnwidth]{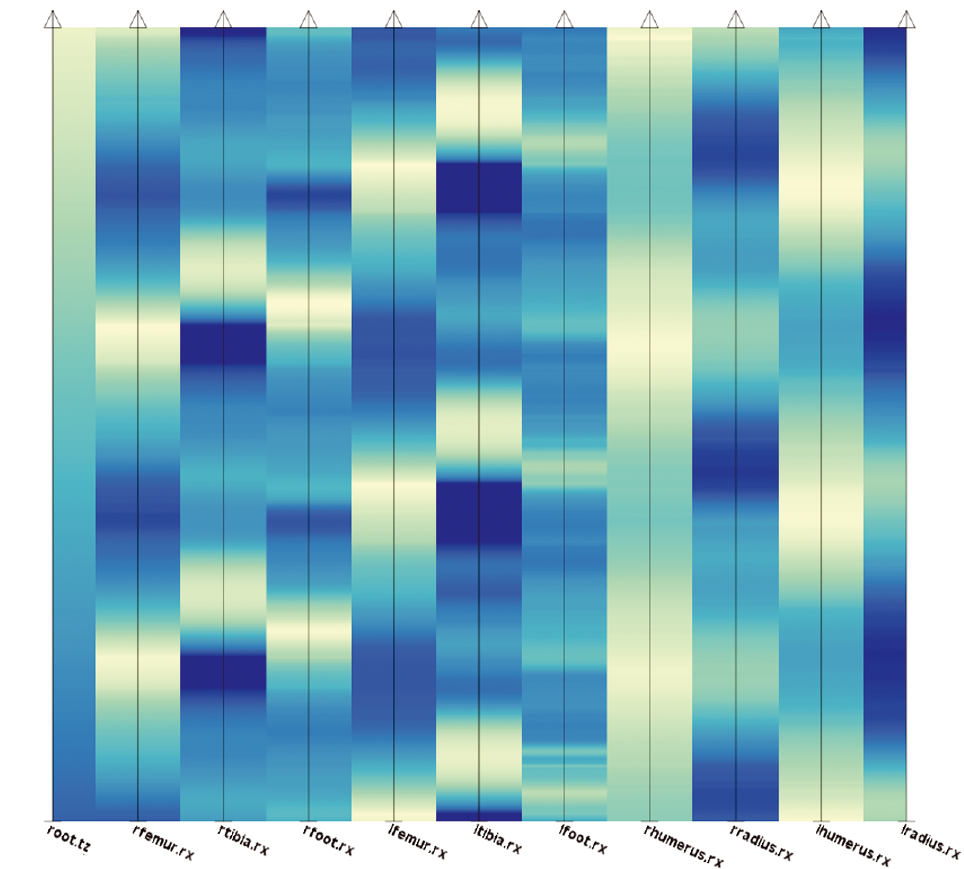}
	\caption{The Temporal heat Map of~\cite{grottel2014visual} showing changes in attribute values over time slices.}
	\label{fig:temporal_heat_map}
\end{figure}

The standard multivariate visualization of parallel coordinates is also a suitable basic visualization technique.
In the case where the graph is multiplex, and nodes appear in all layers, the different axes can represent a specific layer attribute.
Heat maps may also be adapted for a multilayer use case. For example, the temporal heat-maps of Grottel \emph{et al.}~\cite{grottel2014visual} are made suitable for multilayer attribute visualization, by using graph layers instead of time slices for each column (see Figure~\ref{fig:temporal_heat_map}).

An interesting  example of categorical multivariate data in a single layer, which could be extended for multilayer visualization can be seen in the multivariate graph analysis tool of Pretorius and van Wijk~\cite{pretorius2008}. 
Their approach uses icicle plots to describe the (hierarchical) categorical attributes of the source and target of a set of directed edges. 
The source icicle plot is on the left side of the screen and the one for the target nodes is on the right, with the edge and their associated data drawn in the middle. Such an approach may be easily adapted to compare categorical labels across layers Task category~\textbf{B}). 
Combined with edge bundling, as done by Holten~\cite{holten2008} with Hierarchical Edge Bundling, it could also be used to examine structural and categorical attribute difference between layers simultaneously (Task categories~\textbf{B} and~\textbf{D2}).

It is also possible to consider categorical attribute data as a network layer in and of itself. For example, in the application OntoVis~\cite{shen2006}, Shen \emph{et al.} use a node ontology graph to query a large heterogeneous social network dataset. 
The node ontology graph reflects the disparity of the attribute (how well distributed it is across nodes), and its edges display frequency of links between the entities.
It acts both as a visualization of aggregated categorical data, and a layer by which the dataset can be better interacted with and understood.

The approach used for attribute visualization relies heavily on the task the user is performing. For example a scatter-plot matrix is one technique by which attributes may be summarised, possibly even across layers. However if the user's goal is to understand correlations of attributes across layers, an approach such as the modified multilayer version the scatter-plot staircase (SPLOS) of \BPrevised{Viau \etal}~\cite{viau2010} may be more efficient in terms of comprehension and space.
In this approach scatter-plots of the attributes are ordered pairwise based on correlation and common axes.
 
Attribute visualization also can be combined with interaction withing the context of multilayer graph visualization, to help better understand the connection between layers. The Detangler application~\cite{renoust2015detangler} visualizes the level of \emph{entanglement} of a selected set of nodes by colouring the selection lasso (an attribute measuring internal cohesion of a group -- as opposed to group inertia or entropy~\cite{Shannon1948}, also proposed in~\cite{Battiston2014}).

\begin{figure}[!ht]
	\centering
	\includegraphics[width=\columnwidth]{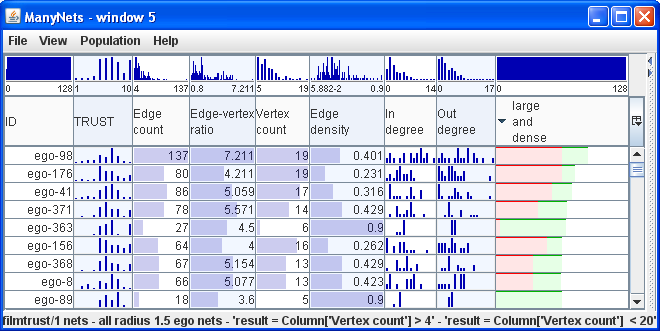}
	\caption{The list view of the Manynets application~\cite{freire2010ManyNets}, summarising attributes of networks using bar charts. The vertical barcharts show the distribution of attribute values and the green and red stacked horizontal bar is a combined score based on several inputs.}
	\label{fig:manynets}
\end{figure}

Attributes should not be considered only at a per node level. Aggregation is an important feature of \emph{GraphTrail}~\cite{dunne2012graphtrail} an application which focuses on exploring multivariate heterogeneous networks. It eschews standard network visualization encodings, such as node-link and matrix, in favour of aggregate attribute visualizations using a hybrid approach bar charts combined with arc diagrams. 
Such an approach is beneficial to the characterisation and understanding of layers and their interactions.
Barchart visualizations are also used by \emph{ManyNets}~\cite{freire2010ManyNets} as a means of summarising and comparison of networks, see Figure~\ref{fig:manynets}. The set of charts describing a network are referred to as ``network -fingerprints'' and the tabular presentation allows for easy comparison and sorting across networks (or layers, depending on the nomenclature chosen).

\subsection{Empirical Evaluation}
\label{sec:evaluation}
Many of the multilayer network visualization papers from the information visualization domain described here are either system papers \eg~\cite{renoust2015detangler,kairam2015refinery} or  design study \eg~\cite{Ghani:2013fo}. Evaluation frequently involves user feedback~\cite{Ghani:2013fo,dunne2012graphtrail}, visualization expert  review~\cite{shi2014Hierarchical}, usage scenarios \cite{dunne2012graphtrail,renoust2015detangler}.
There is a dearth of low level empirical evaluations specific to multilayer network visualizations, although this is partially because there are few clearly low level tasks defined, and there is also a lack of existing techniques to compare them with. For example a novel interaction like that of \emph{Detangler}~\cite{renoust2015detangler} cannot be compared directly to any other technique. Therefore an empirical comparison of user performance at a related low level task is simply not practical.
Within domains external to those related to information visualization there is less demand for performing a thorough evaluation of systems or techniques, so authors may just demonstrate the techniques with a suitable dataset, \eg~\cite{crnovrsanin2014visualization}.
The \emph{MuxViz} application displays layers in a 2D planes in a 3D visualization (a.k.a 2.5D), in one of the many types of visualization it offers. However, as mentioned earlier, no empirical evaluation has ever shown such use of 3D graph visualization to be beneficial, with the exception of when viewed with stereo and/or motion or depth cues~\cite{Ware2008,greffard2011visual}.


%% file: 4_Discussion.tex
\section{Discussion and Open Challenges}\label{sect:discussion}
The goal of this report is a review of a large set of tools and techniques to support the increasing demand for the visualization of multilayer networks. Many of the interesting ideas come from related concepts, such as multivariate and faceted visualization, however neither of these concepts fully encompasses the multilayer network model. 
The existing techniques provide a starting point, however, as a result of the complexity of the systems modelled as multilayer networks, there are still many novel tasks that need to be addressed (Section~\ref{subsec:discussion_taxonomy}), possible improvements for modelling layers (Section~\ref{sub:datadef}), visualization and interaction gaps that need to be filled (Sections~\ref{sub:visapp} to~\ref{sub:attvis}), and empirical user-evaluations to be made (Section~\ref{sub:eval}).


\subsection{Multilayer Networks Task Taxonomy}
\label{subsec:discussion_taxonomy}
Tasks are a motivating force for multilayer network visualization as a topic. There are many existing task taxonomies that cover network visualization as discussed in Section~\ref{subsec:task_analysis}. 
Our taxonomy of tasks extends these existing taxonomies.
The taxonomy of Lee \emph{et al.}~\cite{lee2006task} considers graph comparison as a high level task not covered by their taxonomy. In the definition of multilayer networks, layers become an integral part of the structure and as a result layer related tasks can no longer be considered abstract or high level. They are as fundamental part of a graph task taxonomy as nodes and edges.
However, these aspects of Lee \emph{et al.}'s taxonomy can be applied to the graph entities within each layer.

\subsection{Data Definition}
\label{sub:datadef}
As mentioned in Section~\ref{subsec:data_definition}, many of the approaches, particularly form the infovis domain, did not explicitly mention that the data was a multilayer network. An important part of understanding the data is determining what aspects (and hence layers) need to be visualized to support the users goals as early in the design process as possible.
As described in Section~\ref{subsec:layer_related_concepts}, layers can be considered a characteristic of the multilayer system as a whole, defined either by a physical reality or the system being modelled. However there are still multiple ways to determine the set of layers for analysis. 
\paragraph*{Modelling of real world concepts from the data}
Real systems often begin with raw data and not a graph. However, in many of the papers we have reviewed the systems are presented with fully organised and cleaned data sets, \eg~\cite{kairam2015refinery,shi2014Hierarchical}.
Within the application domains generating a multilayer data set for analysis is often a significant focus of the work \cite{ducruet2017Maritime,Zeng:2016ez,Barthelemey15} independent of visualization.
It is already recognised that creating a general purpose graph from real data is a challenge~\cite{kandel2011research,srinivasan2018grpahiti}, and doing so across multiple layers can be considered even more challenging. 
Existing approaches~\cite{heer2014orion,srinivasan2018grpahiti} consider the problem from a general graph point of view and could be developed further to consider graph aspects and layer definition. 
\paragraph*{Entities that encode layer definitions}
When modelling layers it is easy to consider a node type attribute to characterise an aspect and encode data into layers. However, it is worth emphasising that there are many other options. Multiple aspects can be combined together, \eg in the biological domain one aspect could be omics level and another could be species, resulting in layers that describe an omics level for a specific species. 
Edge types are used in many cases to generate layers (usually in multiplex cases such as~\cite{renoust2015detangler,ducruet2017Maritime}).
It is worth remembering the advice of Kivel\"{a} \emph{et al.}, and be ``creative''. 
\paragraph*{Analytical generation of layers}
The raw data may not map to the real world concepts embodied in a system and may require some processing. If layers are not immediately forthcoming, a \textbf{clustering} approach might reveal structure not explicitly encoded in the data. Consider the example of a predator-prey network, a topological clustering may group animals based on geography even if geography is not explicitly encoded in the data. While the process is analytical, it still results in a layering that reflects the reality of the system being modelled.
\textbf{Projection} is another means by which layers can be created. Bipartite systems can be analysed by projecting on a node type\cite{latapy2008basic}. For example,  a bipartite author-paper network, where researchers are connected to papers that they authored. A projection on the paper node type results in a co-authorship network of researchers, where two researchers are connected if they ever authored a paper.
Such an operation may be adapted to a multilayer user case.
\textbf{Degree of interest (DOI)} functions suggest nodes for inclusion based on what the user has already characterised as interesting. This approach has already been used by the Refinery application and may also be applied to datasets that are explicitly multilayer\cite{laumond2017edoi}.

\subsection{Visualization approaches}
\label{sub:visapp}
As seen in Section~\ref{subsec:survey_visualization} there are a wide range of existing visualization techniques which can used for, or adapted to, visualizing multilayer networks.
There are many aspects of multilayer network visualization that are opportunities for immediate investigation with respect to visualization.

\textbf{Hybrid visualization}, as discussed in Section~\ref{subsec:survey_visualization_hybrid} hybrid visualizations are techniques which can be exploited for multilayer network visualization. Only a small subset of the range of approaches discussed throughout Section~\ref{subsec:survey_visualization} have been combined and hybridized, meaning there are many potential options still to be investigated to support multilayer tasks.

The need to address tasks related to cross layer entity comparison also means that there may be interesting opportunities with respect to \textbf{edge routing and visualization}. The approach used by \cite{crnovrsanin2014visualization} is not developed much beyond the original edge bundling algorithm of Holten and van Wijk, while the  bundling of~\cite{bourqui2016multilayer} focuses on edge routing in the case where the nodes and edges of all layers are presented in a single node-link diagram. 

Within this report we have intentionally avoided focusing on more complex data modelling approaches such as hyper-graphs. However, it is worth noting that in many applications,  especially in the domain of biology, the datasets are explicitly modelled as hyper-graphs, \eg the Systems Biology Graph Notation (SBGN)~\cite{lenovere2009systems} that is often used to describe biological pathways. Representing hyper-edges in a multilayer context (particularly if endpoints belong to discrete layers), is an interesting open challenge.

Some multilayer datasets also contain a temporal aspect, \eg~\cite{Barthelemey15}, and there has been much work done in the field of complex systems on the dynamics of multilayer networks~\cite{boccaletti2014structure}. However integration between temporal and other aspects for dynamic multilayer networks may still offer opportunities for novel visualization techniques.

\subsection{Interaction Approaches}
\label{sub:interacgap}
Multilayer Network related  tasks and exploration may require novel interaction techniques. As described in Section~\ref{subsec:interaction}, \emph{Detangler} is one example of an interaction technique to support multilayer network exploration, supporting Task categories~\textbf{A},~\textbf{C} and~\textbf{D} of our taxonomy. The \emph{Donatien} application of~\cite{Hascoet:2012:IGM:2254556.2254654} supports interaction techniques related to comparison of multiple layers (Task category~\textbf{D2} in our taxonomy), and defining layers for comparison (Task category~\textbf{C}). However, there is still a large design space to be explored for multilayer use cases, particularly inter-layer exploration (Task category~\textbf{D}) and layer creation / manipulation (Task category~\textbf{C}).

\subsection{Attribute visualization}
\label{sub:attvis}
Attribute visualization is important for understanding the differences in attribute values for the same node in different layers, and understand differences at the layer level via aggregation or summarization. However many existing techniques can be adapted relatively easily to the multilayer case, as seen in Section~\ref{subsec:attrib_vis}. 
The most novel attribute visualization, seen in the \emph{Detangler}~\cite{renoust2015detangler} system, is related to a multilayer interaction technique that uses a multilayer metric.
Many classical network centralities have been adapted for the multilayer network use case~\cite{domenico2013centrality,MultilayerNetworks}. While MuxViz~\cite{dedominico2015MuxViz} does include some  visualization of these types of attributes, as shown in Figure~\ref{fig:muxviz_rings}, there is much opportunity for novel attribute visualization considering multilayer centralities, integrated into network visualizations, to support cross layer comparisons incorporating both attributes and structure (Task categories~\textbf{D1} and~\textbf{D2}).

\subsection{Evaluation}
\label{sub:eval}
Task taxonomies are widely accepted to be useful for the evaluation process~\cite{Kerracher2017} and the tasks describes in Section~\ref{subsec:task_analysis}, should support the evaluation of multilayer visualization systems and techniques. As described in Section~\ref{sec:evaluation}, there is a lack of empirical evaluation for multilayer network visualizations. Crowdsourcing offers a lot of promise for information visualization~\cite{borgo2017crowdsourcing}, particularly for evaluation. A survey of evaluation using crowdsourcing in information visualization has shown that while the  tasks for crowdsourcing based evaluations are in the majority of the cases simple tasks~\cite{borgo2018Crowdsourcing}, more complex (and synoptic) tasks are possible. Many existing crowdsourcing platforms do not lend themselves to tasks that are highly interactive, however the development of new platforms driven by academic needs, such as suggested by~\cite{hirth2017CStech}, may simplify evaluating more complex tasks. Crowdsourcing may prove be useful to address the lack of evaluation for approaches to multilayer network visualizations, but the complexity of the tasks and the datasets, for the moment,  makes it challenging.

%% file: 5_Conclusions_End.tex
\section{Conclusion and Roadmap for Future Research}\label{sect:conclusion}
With this paper we have presented a survey showing the state of the art of visualization of multilayer networks within both
the domain of visualization, and others.
We have shown that multilayer network problems are at the intersection of domain and data. There are many existing techniques that address many aspects of multilayer network visualization that may be used in many situations.

We have also identified aspects that require further research. We have identified categories of tasks, not covered by existing network task taxonomies, and have identified immediate opportunities for research on multilayer network visualization. 
We believe that the visualization of multilayer networks will play an important role in the future of network visualization and by working closely with the field of complex systems and the application domains we can uncover, and find solutions, to many new visualization related challenges.  As the field of complex networks grows, more application domains will take advantage of the ability to better model and handle the complexity inherent in the systems being studied.  Bringing the visualization community closer to the application domains communities, as well as the complex systems communities, will result in improved outcomes for all involved. Organising workshops and seminars that include representatives from all communities will help to achieve this goal.
 As they do, they will encounter new and interesting challenges and will need novel visualization (and visual analytics) approaches to address these problems. In our opinion, the roadmap for future research starts by:

\paragraph*{Re-frame user needs and data as multilayer network problems.}
Kivel\"{a} \emph{et al.} discuss the range of data definitions (heterogeneous, multiplex, etc.) that are covered by their framework.  Re-framing a user's problem with these descriptions may prevent commonalities between problems being obscured by nomenclature, but more importantly it will give the visualization researchers more exposure to application domain researchers addressing multilayer network problems.

\paragraph*{Closer interaction with the applications domain communities}
Consolidating and refining multilayer network tasks with the typology of Munzner and Brehmer~\cite{brehmer2013multi} and developing higher level task descriptions with the domains will allow for a better understanding of both the core elements of problems across domains and the full range of solutions available.

\paragraph*{Closer interaction with the complex systems community}
To better understand the data, closer interaction with the complex systems community will allow for the use of novel analytic approaches. Multilayer analytics have not been fully exploited in support of visualization, and we have only touched on a few key aspects in this survey. There is a vast amount of new multilayer network analytics which may be part of the answer to the visualization challenges that arise from the application domains.

\subsection*{Acknowledgements}
This work was (partially) funded by the French ANR grant BLIZAAR ANR-15-CE23-0002-01 and the Luxembourgish FNR grant BLIZAAR INTER/ANR/14/9909176.

%% file: main.bbl
\begin{thebibliography}{100}

\bibitem{Agarwal2017}
Shivam Agarwal, Amit Tomar, and Jaya Sreevalsan-Nair.
\newblock Nodetrix-multiplex: Visual analytics of multiplex small world
  networks.
\newblock In Hocine Cherifi, Sabrina Gaito, Walter Quattrociocchi, and
  Alessandra Sala, editors, {\em Complex Networks \& Their Applications V},
  volume 693 of {\em Studies in Computational intelligence}, pages 579--591,
  \BPrevised{Milano, Italy}, 2017. Springer.

\bibitem{Ahn2013Task}
Jae-wook Ahn, Catherine Plaisant, and Ben Shneiderman.
\newblock A task taxonomy for network evolution analysis.
\newblock {\em IEEE Transactions on Visualization and Computer Graphics},
  99(PP):365--376, 2013.

\bibitem{alsallakh2013radial}
Bilal Alsallakh, Wolfgang Aigner, Silvia Miksch, and Helwig Hauser.
\newblock Radial sets: Interactive visual analysis of large overlapping sets.
\newblock {\em IEEE Transactions on Visualization and Computer Graphics},
  19(12):2496--2505, 2013.

\bibitem{andrienko2006exploratory}
Natalia Andrienko and Gennady Andrienko.
\newblock {\em Exploratory analysis of spatial and temporal data: a systematic
  approach}.
\newblock \BPrevised{Springer, Berlin, Heidelberg}, 2006.

\bibitem{Atzmueller2012}
Martin Atzmueller, Stephan Doerfel, Andreas Hotho, Folke Mitzlaff, and Gerd
  Stumme.
\newblock Face-to-face contacts at a conference: dynamics of communities and
  roles.
\newblock In Martin Atzmueller, Alvin Chin, Denis Helic, and Andreas Hotho,
  editors, {\em Modeling and Mining Ubiquitous Social Media \BPrevised{(MSM
  2011)}}, volume 7472 of {\em Lecture Notes in Computer Science}, pages
  21--39, \BPrevised{Athens, Greece}, 2012. Springer Berlin Heidelberg.

\bibitem{Battiston2014}
Federico Battiston, Vincenzo Nicosia, and Vito Latora.
\newblock Structural measures for multiplex networks.
\newblock {\em Physical Review E}, 89(3):032804, 2014.

\bibitem{beck2014state}
Fabian Beck, Michael Burch, Stephan Diehl, and Daniel Weiskopf.
\newblock A taxonomy and survey of dynamic graph visualization.
\newblock {\em Computer Graphics Forum}, 36(1):133--159, 2017.

\bibitem{Bentley2016}
Barry Bentley, Robyn Branicky, Christopher~L. Barnes, Yee~Lian Chew, Eviatar
  Yemini, Edward~T. Bullmore, Petra~E. V\'{e}rtes, and William~R. Schafer.
\newblock The multilayer connectome of caenorhabditis elegans.
\newblock {\em PLOS Computational Biology}, 12(12):1--31, 2016.

\bibitem{Bezerianos:2010vb}
Anastasia Bezerianos, Fanny Chevalier, Pierre Dragicevic, Niklas Elmqvist, and
  Jean-Daniel Fekete.
\newblock Graphdice: A system for exploring multivariate social networks.
\newblock {\em Computer Graphics Forum}, 29(3):863--872, 2010.

\bibitem{boccaletti2014structure}
Stefano Boccaletti, Ginestra Bianconi, Regino Criado, Charo~I Del~Genio,
  Jes{\'u}s G{\'o}mez-Gardenes, Miguel Romance, Irene Sendina-Nadal, Zhen Wang,
  and Massimiliano Zanin.
\newblock The structure and dynamics of multilayer networks.
\newblock {\em Physics Reports}, 544(1):1--122, 2014.

\bibitem{borgatti1997network}
Stephen~P Borgatti and Martin~G Everett.
\newblock Network analysis of 2-mode data.
\newblock {\em Social networks}, 19(3):243--269, 1997.

\bibitem{borgo2017crowdsourcing}
Rita Borgo, Bongshin Lee, Benjamin Bach, Sara Fabrikant, Radu Jianu, Andreas
  Kerren, Stephen Kobourov, Fintan McGee, Luana Micallef, Tatiana von
  Landesberger, et~al.
\newblock Crowdsourcing for information visualization: Promises and pitfalls.
\newblock In {\em Evaluation in the Crowd. Crowdsourcing and Human-Centered
  Experiments}, volume 10264 of {\em Lecture Notes in Computer Science}, pages
  96--138. Springer, 2017.

\bibitem{borgo2018Crowdsourcing}
Rita Borgo, Luana Micallef, Benjamin Bach, Fintan McGee, and Bongshin Lee.
\newblock Information visualization evaluation using crowdsourcing.
\newblock {\em Computer Graphics Forum}, 37(3):573--595, 2018.

\bibitem{bothorel2013visualization}
Gwenael Bothorel, Mathieu Serrurier, and Christophe Hurter.
\newblock Visualization of frequent itemsets with nested circular layout and
  bundling algorithm.
\newblock In {\em International Symposium on Visual Computing}, volume 8034 of
  {\em Lecture Notes in Computer Science}, pages 396--405,
  \BPrevised{Rethymnon, Crete, Greece}, 2013. Springer.

\bibitem{bourqui2016multilayer}
R.~Bourqui, D.~Ienco, A.~Sallaberry, and P.~Poncelet.
\newblock Multilayer graph edge bundling.
\newblock In {\em IEEE Pacific Visualization Symposium (PacificVis)}, pages
  184--188, \BPrevised{Taipei, Taiwan}, April 2016.

\bibitem{brehmer2013multi}
Matthew Brehmer and Tamara Munzner.
\newblock A multi-level typology of abstract visualization tasks.
\newblock {\em IEEE Transactions on Visualization and Computer Graphics},
  19(12):2376--2385, 2013.

\bibitem{Bright2015}
David~A. Bright, Catherine Greenhill, Alison Ritter, and Carlo Morselli.
\newblock Networks within networks: using multiple link types to examine
  network structure and identify key actors in a drug trafficking operation.
\newblock {\em Global Crime}, 16(3):219--237, 2015.

\bibitem{buldyrev2010catastrophic}
Sergey~V Buldyrev, Roni Parshani, Gerald Paul, H~Eugene Stanley, and Shlomo
  Havlin.
\newblock Catastrophic cascade of failures in interdependent networks.
\newblock {\em Nature}, 464:1025--1028, 2010.

\bibitem{Burt1985}
Ronald Burt and Thomas Sch{\o}tt.
\newblock Relation content in multiple networks.
\newblock {\em Social Science Research}, 14(4):287--308, 1985.

\bibitem{Cao:2015dy}
Nan Cao, Yu-Ru Lin, Liangyue Li, and Hanghang Tong.
\newblock g-miner: interactive visual group mining on multivariate graphs.
\newblock In {\em \BPrevised{Proceedings of the $33^{rd}$ Annual ACM Conference
  on Human Factors in Computing Systems}}, pages 279--288, \BPrevised{Seoul,
  Republic of Korea}, 2015. ACM.

\bibitem{cao2010facetatlas}
Nan Cao, Jimeng Sun, Yu-Ru Lin, David Gotz, Shixia Liu, and Huamin Qu.
\newblock Facetatlas: Multifaceted visualization for rich text corpora.
\newblock {\em IEEE transactions on visualization and computer graphics},
  16(6):1172--1181, 2010.

\bibitem{cardillo2013emergence}
Alessio Cardillo, Jes{\'{u}}s G{\'{o}}mez-Garde{\~{n}}es, Massimiliano Zanin,
  Miguel Romance, David Papo, Francisco del Pozo, and Stefano Boccaletti.
\newblock Emergence of network features from multiplexity.
\newblock {\em Nature}, 3:1344, feb 2013.

\bibitem{chen2002generalized}
Chun-Houh Chen.
\newblock Generalized association plots: Information visualization via
  iteratively generated correlation matrices.
\newblock {\em Statistica Sinica}, 12(1):7--29, 2002.

\bibitem{Chuang:2012:TVT:2254556.2254572}
Jason Chuang, Christopher~D. Manning, and Jeffrey Heer.
\newblock Termite: Visualization techniques for assessing textual topic models.
\newblock In {\em Proceedings of the International Working Conference on
  Advanced Visual Interfaces}, AVI '12, pages 74--77, \BPrevised{Capri Island,
  Italy}, 2012. ACM.

\bibitem{collins2007vislink}
Christopher Collins and Sheelagh Carpendale.
\newblock Vislink: Revealing relationships amongst visualizations.
\newblock {\em IEEE Transactions on Visualization and Computer Graphics},
  13(6):1192--1199, 2007.

\bibitem{cottret2010}
Ludovic Cottret, David Wildridge, Florence Vinson, Michael~P. Barrett, Hubert
  Charles, Marie-France Sagot, and Fabien Jourdan.
\newblock Metexplore: a web server to link metabolomic experiments and
  genome-scale metabolic networks.
\newblock {\em Nucleic Acids Research}, 38(Web server issue):W132--W137, 2010.

\bibitem{crabb2017disease}
Helen~Kathleen Crabb, Joanne~Lee Allen, Joanne~Maree Devlin, Simon~Matthew
  Firestone, Mark~Anthony Stevenson, and James~Rudkin Gilkerson.
\newblock The use of social network analysis to examine the transmission of
  salmonella spp. within a vertically integrated broiler enterprise.
\newblock {\em Food Microbiology}, 71:73--81, 2017.

\bibitem{crnovrsanin2014visualization}
Tarik Crnovrsanin, Chris~W. Muelder, Robert Faris, Diane Felmlee, and Kwan-Liu
  Ma.
\newblock Visualization techniques for categorical analysis of social networks
  with multiple edge sets.
\newblock {\em Social Networks}, 37(Supplement C):56 -- 64, 2014.

\bibitem{deDomenico2017}
Manlio De~Domenico.
\newblock Multilayer modeling and analysis of human brain networks.
\newblock {\em GigaScience}, 6(5):1--8, 2017.

\bibitem{dedominico2015MuxViz}
Manlio De~Domenico, Mason~A. Porter, and Alex Arenas.
\newblock Muxviz: a tool for multilayer analysis and visualization of networks.
\newblock {\em Journal of Complex Networks}, 3(2):159--176, 2015.

\bibitem{derrible2017complexity}
Sybil Derrible.
\newblock Complexity in future cities: the rise of networked infrastructure.
\newblock {\em International Journal of Urban Sciences}, 21(sup1):68--86, 2017.

\bibitem{Dickison2016}
Mark~E. Dickison, Matteo Magnani, and Luca Rossi.
\newblock {\em Multilayer Social Networks}.
\newblock Cambridge University Press, 2016.

\bibitem{dickison_magnani_rossi_2016}
Mark~E. Dickison, Matteo Magnani, and Luca Rossi.
\newblock {\em Visualizing Multilayer Networks}, pages 79--95.
\newblock Cambridge University Press, 2016.

\bibitem{domenico2013centrality}
MD~Domenico, A~Sol-Ribalta, E~Omodei, S~Gmez, and A~Arenas.
\newblock Centrality in interconnected multilayer networks.
\newblock {\em Nature Communications}, 6(6868), 2013.

\bibitem{ducruet2017Maritime}
César Ducruet.
\newblock Multilayer dynamics of complex spatial networks: The case of global
  maritime flows (1977--2008).
\newblock {\em Journal of Transport Geography}, 60:47--58, 2017.

\bibitem{DUNBAR2015}
R.I.M. Dunbar, Valerio Arnaboldi, Marco Conti, and Andrea Passarella.
\newblock The structure of online social networks mirrors those in the offline
  world.
\newblock {\em Social Networks}, 43:39 -- 47, 2015.

\bibitem{dunne2012graphtrail}
Cody Dunne, Nathalie Henry~Riche, Bongshin Lee, Ronald Metoyer, and George
  Robertson.
\newblock Graphtrail: Analyzing large multivariate, heterogeneous networks
  while supporting exploration history.
\newblock In {\em Proceedings of the SIGCHI Conference on Human Factors in
  Computing Systems}, CHI '12, pages 1663--1672, \BPrevised{Austin, Texas,
  USA}, 2012. ACM.

\bibitem{engle2012visualizing}
Sophie Engle and Sean Whalen.
\newblock Visualizing distributed memory computations with hive plots.
\newblock In {\em Proceedings of the Ninth International Symposium on
  Visualization for Cyber Security (VizSec'12)}, pages 56--63,
  \BPrevised{Seattle, WA, USA}, 2012. ACM.

\bibitem{fekete2015reorder}
Jean-Daniel Fekete.
\newblock Reorder.js: A javascript library to reorder tables and networks.
\newblock In {\em VIS 2015 Poster}, \BPrevised{Chicago, USA}, 2015. IEEE.

\bibitem{freire2010ManyNets}
Manuel Freire, Catherine Plaisant, Ben Shneiderman, and Jen Golbeck.
\newblock Manynets: An interface for multiple network analysis and
  visualization.
\newblock In {\em Proceedings of the SIGCHI Conference on Human Factors in
  Computing Systems}, CHI '10, pages 213--222, \BPrevised{Atlanta, Georgia,
  USA}, 2010. ACM.

\bibitem{fung2009visual}
David~CY Fung, Seok-Hee Hong, Dirk Kosch{\"u}tzki, Falk Schreiber, and Kai Xu.
\newblock Visual analysis of overlapping biological networks.
\newblock In {\em $13^{th}$ International Conference Information
  Visualisation}, pages 337--342, \BPrevised{Barcelona, Spain}, 2009. IEEE.

\bibitem{Barthelemey15}
Riccardo Gallotti and Marc Barthelemy.
\newblock The multilayer temporal network of public transport in great britain.
\newblock {\em Nature Scientific Data}, 2(140056), 2015.

\bibitem{gao2012networks}
Jianxi Gao, Sergey~V. Buldyrev, H.~Eugene Stanley, and Shlomo Havlin.
\newblock Networks formed from interdependent networks.
\newblock {\em Nature physics}, 8(1):40--48, 2012.

\bibitem{Geard2007}
N.L. Geard and Seth Bullock.
\newblock Milieu and function: Toward a multilayer framework for understanding
  social networks.
\newblock In {\em Workshop Proceedings of the Ninth European Conference on
  Artificial Life (ECAL): The Emergence of Social Behaviour}, pages 1--11,
  \BPrevised{Portugal}, 2007.

\bibitem{gehlenborg2010visualization}
Nils Gehlenborg, Se{\'a}n~I O'donoghue, Nitin~S Baliga, Alexander Goesmann,
  Matthew~A Hibbs, Hiroaki Kitano, Oliver Kohlbacher, Heiko Neuweger, Reinhard
  Schneider, Dan Tenenbaum, et~al.
\newblock Visualization of omics data for systems biology.
\newblock {\em Nature methods}, 7(3 Suppl.):S56--68, 2010.

\bibitem{Ghani:2013fo}
S~Ghani, Bum~Chul Kwon, Seungyoon Lee, Ji~Soo Yi, and N~Elmqvist.
\newblock Visual analytics for multimodal social network analysis: A design
  study with social scientists.
\newblock {\em Transactions on Visualization and Computer Graphics},
  19(12):2032--2041, 2013.

\bibitem{ghoniem2005peeking}
Mohammad Ghoniem, Hadrien Cambazard, Jean-Daniel Fekete, and Narendra Jussien.
\newblock Peeking in solver strategies using explanations visualization of
  dynamic graphs for constraint programming.
\newblock In {\em Proceedings of the 2005 ACM symposium on Software
  visualization (SoftVis)}, pages 27--36, \BPrevised{St. Louis, Missouri, USA},
  2005. ACM.

\bibitem{ghoniem2005readability}
Mohammad Ghoniem, Jean-Daniel Fekete, and Philippe Castagliola.
\newblock On the readability of graphs using node-link and matrix-based
  representations: a controlled experiment and statistical analysis.
\newblock {\em Information Visualization}, 4(2):114--135, 2005.

\bibitem{gosak2017}
Marko Gosak, Rene Markovi\v{c}, Jurij Dolen\v{s}ek, Marjan~Slak Rupnik, Marko
  Marhl, Andra\v{z} Sto\v{z}er, and Matja\v{z} Perc.
\newblock Network science of biological systems at different scales: A review.
\newblock {\em Physics of Life Reviews}, 2017.

\bibitem{greffard2011visual}
Nicolas Greffard, Fabien Picarougne, and Pascale Kuntz.
\newblock Visual community detection: An evaluation of {2D}, {3D} perspective
  and {3D} stereoscopic displays.
\newblock In Marc van Kreveld and Bettina Speckmann, editors, {\em Graph
  Drawing}, pages 215--225, \BPrevised{Redmond, WA, USA}, 2012. Springer Berlin
  Heidelberg.

\bibitem{grottel2014visual}
Sebastian Grottel, Julian Heinrich, Daniel Weiskopf, and Stefan Gumhold.
\newblock Visual analysis of trajectories in multi-dimensional state spaces.
\newblock {\em Computer Graphics Forum}, 33(6):310--321, 2014.

\bibitem{hadlak2015survey}
Steffen Hadlak, Heidrun Schumann, and Hans-J{\"o}rg Schulz.
\newblock A survey of multi-faceted graph visualization.
\newblock In {\em Eurographics Conference on Visualization (EuroVis). The
  Eurographics Association}, pages 1--20, \BPrevised{Cagliary, Italy}, 2015.

\bibitem{Halu2014}
Arda Halu, Satyam Mukherjee, and Ginestra Bianconi.
\newblock Emergence of overlap in ensembles of spatial multiplexes and
  statistical mechanics of spatial interacting network ensembles.
\newblock {\em Phys. Rev. E}, 89:012806, Jan 2014.

\bibitem{Hascoet:2012:IGM:2254556.2254654}
Mountaz Hasco\"{e}t and Pierre Dragicevic.
\newblock Interactive graph matching and visual comparison of graphs and
  clustered graphs.
\newblock In {\em Proceedings of the International Working Conference on
  Advanced Visual Interfaces}, AVI '12, pages 522--529, Capri Island, Italy,
  2012. ACM.

\bibitem{heath2009multimodal}
Lenwood~S Heath and Allan~A Sioson.
\newblock Multimodal networks: Structure and operations.
\newblock {\em IEEE/ACM Transactions on Computational Biology and
  Bioinformatics (TCBB)}, 6(2):321--332, 2009.

\bibitem{heer2014orion}
Jeffrey Heer and Adam Perer.
\newblock Orion: A system for modeling, transformation and visualization of
  multidimensional heterogeneous networks.
\newblock {\em Information Visualization}, 13(2):111--133, 2014.

\bibitem{Henry:2007er}
N~Henry, J~D Fekete, and M~J McGuffin.
\newblock Nodetrix: a hybrid visualization of social networks.
\newblock {\em IEEE Transactions on Visualization and Computer Graphics},
  13(6):1302--1309, 2007.

\bibitem{hirth2017CStech}
Matthias Hirth, Jason Jacques, Peter Rodgers, Ognjen Scekic, and Michael
  Wybrow.
\newblock Crowdsourcing technology to support academic research.
\newblock In Daniel Archambault, Helen Purchase, and Tobias Ho{\ss}feld,
  editors, {\em Evaluation in the Crowd. Crowdsourcing and Human-Centered
  Experiments}, volume 10264 of {\em Lecture Notes in Computer Science}, pages
  70--95. Springer International Publishing, 2017.

\bibitem{hoffswell2018setcola}
Jane Hoffswell, Alan Borning, and Jeffrey Heer.
\newblock Setcola: High-level constraints for graph layout.
\newblock {\em Computer Graphics Forum}, 37(3):537--548, 2018.

\bibitem{holten2006}
D.~Holten.
\newblock Hierarchical edge bundles: Visualization of adjacency relations in
  hierarchical data.
\newblock {\em IEEE Transactions on Visualization and Computer Graphics},
  12(5):741--748, Sept 2006.

\bibitem{holten2008}
Danny Holten and Jarke~J. Van~Wijk.
\newblock Visual comparison of hierarchically organized data.
\newblock {\em Computer Graphics Forum}, 27(3):759--766, 2008.

\bibitem{huang2015}
Dandan Huang, Melanie Tory, Bon~Adriel Aseniero, Lyn Bartram, Scott Bateman,
  Sheelagh Carpendale, Anthony Tang, and Robert Woodbury.
\newblock Personal visualization and personal visual analytics.
\newblock {\em Visualization and Computer Graphics, IEEE Transactions on},
  21:420--433, 03 2015.

\bibitem{humayoun2016social}
Shah~Rukh Humayoun, Hafez Ezaiza, Ragaad AlTarawneh, and Achim Ebert.
\newblock Social-circles exploration through interactive multi-layered chord
  layout.
\newblock In {\em Proceedings of the International Working Conference on
  Advanced Visual Interfaces (AVI)}, pages 314--315, \BPrevised{Bari, Italy},
  2016. ACM.

\bibitem{inselberg1990parallel}
A.~Inselberg and B.~Dimsdale.
\newblock Parallel coordinates: a tool for visualizing multi-dimensional
  geometry.
\newblock In {\em Proceedings of the First IEEE Conference on Visualization:
  Visualization `90}, pages 361--378, \BPrevised{San Francisco, CA, USA}, Oct
  1990.

\bibitem{Javed:2012vt}
Waqas Javed and Niklas Elmqvist.
\newblock Exploring the design space of composite visualization.
\newblock In {\em IEEE Pacific Visualization Symposium (PacificVis)}, pages
  1--8, \BPrevised{Songdo, Republic of Korea}, 2012.

\bibitem{kairam2015refinery}
S.~Kairam, N.~H. Riche, S.~Drucker, R.~Fernandez, and J.~Heer.
\newblock Refinery: Visual exploration of large, heterogeneous networks through
  associative browsing.
\newblock {\em Computer Graphics Forum}, 34(3):301--310, 2015.

\bibitem{kandel2011research}
Sean Kandel, Jeffrey Heer, Catherine Plaisant, Jessie Kennedy, Frank van Ham,
  Nathalie~Henry Riche, Chris Weaver, Bongshin Lee, Dominique Brodbeck, and
  Paolo Buono.
\newblock Research directions in data wrangling: Visualizations and
  transformations for usable and credible data.
\newblock {\em Information Visualization}, 10(4):271--288, 2011.

\bibitem{KENETT2015}
Dror~Y. Kenett, Matja\u{z} Perc, and Stefano Boccaletti.
\newblock Networks of networks -- {An Introduction}.
\newblock {\em Chaos, Solitons \& Fractals}, 80:1--6, 2015.

\bibitem{Kerracher2017}
N.~Kerracher and J.~Kennedy.
\newblock Constructing and evaluating visualisation task classifications:
  Process and considerations.
\newblock {\em Computer Graphics Forum}, 36(3):47--59, 2017.

\bibitem{Kerracher:2015ik}
Natalie Kerracher, Jessie Kennedy, and Kevin Chalmers.
\newblock A task taxonomy for temporal graph visualisation.
\newblock {\em IEEE Transactions on Visualization and Computer Graphics},
  PP(99):1--1, 2015.

\bibitem{kerren2014introduction}
Andreas Kerren, Helen~C Purchase, and Matthew~O Ward.
\newblock Introduction to multivariate network visualization.
\newblock In {\em Multivariate Network Visualization}, volume 8380 of {\em
  Lecture Notes in Computer Science}, pages 1--9. Springer, 2014.

\bibitem{Kerren2014}
Andreas Kerren and Falk Schreiber.
\newblock Network visualization for integrative bioinformatics.
\newblock In Ming Chen and Ralf Hofest{\"a}dt, editors, {\em Approaches in
  Integrative Bioinformatics: Towards the Virtual Cell}, pages 173--202.
  Springer, 2014.

\bibitem{MultilayerNetworks}
Mikko Kivel\"a, Alex Arenas, Marc Barthelemy, James~P. Gleeson, Yamir Moreno,
  and Mason~A. Porter.
\newblock Multilayer networks.
\newblock {\em Journal of Complex Networks}, 2(3):203--271, 2014.

\bibitem{kohlbacher2014multivariate}
Oliver Kohlbacher, Falk Schreiber, and Matthew~O Ward.
\newblock Multivariate networks in the life sciences.
\newblock In {\em Multivariate Network Visualization}, volume 8380 of {\em
  Lecture Notes in Computer Science}, pages 61--73. Springer, 2014.

\bibitem{krzywinski2011hive}
Martin Krzywinski, Inanc Birol, Steven~JM Jones, and Marco~A Marra.
\newblock Hive plots--rational approach to visualizing networks.
\newblock {\em Briefings in bioinformatics}, 13(5):627--644, 2011.

\bibitem{krzywinski2009circos}
Martin Krzywinski, Jacqueline Schein, Inanc Birol, Joseph Connors, Randy
  Gascoyne, Doug Horsman, Steven~J Jones, and Marco~A Marra.
\newblock Circos: an information aesthetic for comparative genomics.
\newblock {\em Genome research}, 19(9):1639--1645, 2009.

\bibitem{kuo20133omics}
Tien-Chueh Kuo, Tze-Feng Tian, and Yufeng~Jane Tseng.
\newblock 3omics: a web-based systems biology tool for analysis, integration
  and visualization of human transcriptomic, proteomic and metabolomic data.
\newblock {\em BMC systems biology}, 7:64, 2013.

\bibitem{latapy2008basic}
Matthieu Latapy, Cl{\'e}mence Magnien, and Nathalie Del~Vecchio.
\newblock Basic notions for the analysis of large two-mode networks.
\newblock {\em Social networks}, 30(1):31--48, 2008.

\bibitem{laumond2017edoi}
Antoine Laumond, Guy Melan{\c{c}}on, and Bruno Pinaud.
\newblock edoi: Exploratory degree of interest exploration of multilayer
  networks based on user interest.
\newblock In {\em VIS 2017, Poster session}, \BPrevised{Phoenix, AZ, USA},
  2017.

\bibitem{Lazega1999}
E.~Lazega and P.~E. Pattison.
\newblock Multiplexity, generalized exchange and cooperation in organizations:
  A case study.
\newblock {\em Social Networks}, 21:67--90, 1999.

\bibitem{lenovere2009systems}
Nicolas Le~Novere, Michael Hucka, Huaiyu Mi, Stuart Moodie, Falk Schreiber,
  Anatoly Sorokin, Emek Demir, Katja Wegner, Mirit~I Aladjem, Sarala~M
  Wimalaratne, et~al.
\newblock The systems biology graphical notation.
\newblock {\em Nature biotechnology}, 27(8):735--741, 2009.

\bibitem{lee2006task}
Bongshin Lee, Catherine Plaisant, Cynthia~Sims Parr, Jean-Daniel Fekete, and
  Nathalie Henry.
\newblock Task taxonomy for graph visualization.
\newblock In {\em Proceedings AVI workshop on BEyond time and errors: novel
  evaluation methods for information visualization}, pages 1--5,
  \BPrevised{Venice, Italy}, 2006. ACM.

\bibitem{lee2009facetlens}
Bongshin Lee, Greg Smith, George~G Robertson, Mary Czerwinski, and Desney~S
  Tan.
\newblock Facetlens: exposing trends and relationships to support sensemaking
  within faceted datasets.
\newblock In {\em Proceedings of the SIGCHI Conference on Human Factors in
  Computing Systems}, pages 1293--1302, \BPrevised{Boston, MA, USA}, 2009. ACM.

\bibitem{liiv2010seriation}
Innar Liiv.
\newblock Seriation and matrix reordering methods: An historical overview.
\newblock {\em Statistical Analysis and Data Mining: The ASA Data Science
  Journal}, 3(2):70--91, 2010.

\bibitem{lin2008}
Nan Lin.
\newblock A network theory of social capital.
\newblock In Dario Castiglione, Jan~W. van Deth, , and Guglielmo Wolleb,
  editors, {\em The handbook of social capital}, page~69. Oxford University
  Press, 2008.

\bibitem{Liu:2015bf}
Xiaotong Liu and Han-Wei Shen.
\newblock The effects of representation and juxtaposition on graphical
  perception of matrix visualization.
\newblock In {\em \BPrevised{Proceedings of the $33^{rd}$ Annual ACM Conference
  on Human Factors in Computing Systems }}, pages 269--278, \BPrevised{Seoul,
  Republic of Korea}, 2015. ACM Press.

\bibitem{Liu:2017it}
Yuhua Liu, Changbo Wang, Peng Ye, and Kang Zhang.
\newblock Hybridvis: An adaptive hybrid-scale visualization of multivariate
  graphs.
\newblock {\em Journal of Visual Languages \& Computing}, 41:100--110, 2017.

\bibitem{mackinlay1986}
Jock Mackinlay.
\newblock Automating the design of graphical presentations of relational
  information.
\newblock {\em ACM Transactions On Graphics (Tog)}, 5(2):110--141, 1986.

\bibitem{mcgee2016towards}
Fintan McGee, Marten During, and Mohammad Ghoniem.
\newblock Towards visual analytics of multilayer graphs for digital cultural
  heritage.
\newblock In {\em $1^{st}$ Workshop on Visualization for the Digital
  Humanities}, Baltimore, USA, 2016.

\bibitem{moody2005dynamic}
James Moody, Daniel McFarland, and Skye Bender-deMoll.
\newblock Dynamic network visualization.
\newblock {\em American Journal of Sociology}, 110(4):1206--1241, 2005.

\bibitem{moreno1953}
J.L. Moreno.
\newblock {\em Who Shall Survive?}
\newblock Beacon House Inc., 2nd edition, 1953.

\bibitem{Munzner2009}
Tamara Munzner.
\newblock A nested process model for visualization design and validation.
\newblock {\em IEEE Transactions on Visualization and Computer Graphics},
  15(6):921--928, 2009.

\bibitem{Murray2017Taxonomy}
Paul Murray, Fintan McGee, and Angus~G. Forbes.
\newblock A taxonomy of visualization tasks for the analysis of biological
  pathway data.
\newblock {\em BMC Bioinformatics}, 18(2):21, Feb 2017.

\bibitem{pastor2015epidemic}
Romualdo Pastor-Satorras, Claudio Castellano, Piet Van~Mieghem, and Alessandro
  Vespignani.
\newblock Epidemic processes in complex networks.
\newblock {\em Reviews of modern physics}, 87(3):925, 2015.

\bibitem{pavlopoulos2008arena3d}
Georgios~A Pavlopoulos, Se{\'a}n~I O'Donoghue, Venkata~P Satagopam, Theodoros~G
  Soldatos, Evangelos Pafilis, and Reinhard Schneider.
\newblock {Arena3D}: visualization of biological networks in {3D}.
\newblock {\em BMC systems biology}, 2:104, 2008.

\bibitem{Plaisant2004Evaluation}
Catherine Plaisant.
\newblock The challenge of information visualization evaluation.
\newblock In {\em Proceedings of the Working Conference on Advanced Visual
  Interfaces}, AVI '04, pages 109--116, Gallipoli, Italy, 2004. ACM.

\bibitem{pretorius2008}
A.~Johannes Pretorius and Jarke~J. Van~Wijk.
\newblock Visual inspection of multivariate graphs.
\newblock {\em Computer Graphics Forum}, 27(3):967--974, 2008.

\bibitem{pretorius2014tasks}
Johannes Pretorius, Helen~C Purchase, and John~T Stasko.
\newblock {\em Tasks for multivariate network analysis}, pages 77--95.
\newblock Springer, 2014.

\bibitem{purchase2012experimental}
Helen~C Purchase.
\newblock {\em Experimental human-computer interaction: a practical guide with
  visual examples}.
\newblock Cambridge University Press, 2012.

\bibitem{reis2014}
Saulo D.~S. Reis, Yanqing Hu, Andr{\'{e}}s Babino, Jos{\'{e}}~S. Andrade,
  Santiago Canals, Mariano Sigman, and Hern{\'{a}n}~A. Makse.
\newblock Avoiding catastrophic failure in correlated networks of networks.
\newblock {\em Nature Physics}, 10(10):762--767, sep 2014.

\bibitem{ren2018generating}
H.~Ren, B.~Renoust, M.~Viaud, G.~Melançon, and S.~Satoh.
\newblock Generating “visual clouds” from multiplex networks for tv news
  archive query visualization.
\newblock In {\em 2018 International Conference on Content-Based Multimedia
  Indexing (CBMI)}, pages 1--6, Sept 2018.

\bibitem{renoust2015detangler}
B.~Renoust, G.~Melan{\c{c}}on, and T.~Munzner.
\newblock Detangler: Visual analytics for multiplex networks.
\newblock {\em Computer Graphics Forum}, 34(3):321--330, 2015.

\bibitem{Renoust2014}
Benjamin Renoust, Guy Melançon, and Marie-Luce Viaud.
\newblock Entanglement in multiplex networks: Understanding group cohesion in
  homophily networks.
\newblock In Rokia Missaoui and Idrissa Sarr, editors, {\em Social Network
  Analysis - Community Detection and Evolution}, Lecture Notes in Social
  Networks, pages 89--117. Springer, 2014.

\bibitem{rossi2015}
Luca Rossi and Matteo Magnani.
\newblock Towards effective visual analytics on multiplex and multilayer
  networks.
\newblock {\em Chaos, Solitons \& Fractals}, 72:68 -- 76, 2015.
\newblock Multiplex Networks: Structure, Dynamics and Applications.

\bibitem{rufiange2012treematrix}
Sébastien Rufiange, Michael~J. McGuffin, and Christopher~P. Fuhrman.
\newblock Treematrix: A hybrid visualization of compound graphs.
\newblock {\em Computer Graphics Forum}, 31(1):89--101, 2012.

\bibitem{santamaria2008visual}
Rodrigo Santamar{\'\i}a, Roberto Ther{\'o}n, and Luis Quintales.
\newblock A visual analytics approach for understanding biclustering results
  from microarray data.
\newblock {\em BMC bioinformatics}, 9:247, 2008.

\bibitem{saumell2012epidemic}
Anna Saumell-Mendiola, M~{\'A}ngeles Serrano, and Mari{\'a}n Bogun{\'a}.
\newblock Epidemic spreading on interconnected networks.
\newblock {\em Physical Review E}, 86(2), 2012.

\bibitem{schreiber2014heterogeneous}
Falk Schreiber, Andreas Kerren, Katy B{\"o}rner, Hans Hagen, and Dirk Zeckzer.
\newblock Heterogeneous networks on multiple levels.
\newblock In Andreas Kerren, Helen~C. Purchase, and Matthew~O. Ward, editors,
  {\em Multivariate Network Visualization: Dagstuhl Seminar {\#}13201, Dagstuhl
  Castle, Germany, May 12-17, 2013, Revised Discussions}, pages 175--206.
  Springer International Publishing, 2014.

\bibitem{Shadoan2013}
R.~Shadoan and C.~Weaver.
\newblock Visual analysis of higher-order conjunctive relationships in
  multidimensional data using a hypergraph query system.
\newblock {\em IEEE Transactions on Visualization and Computer Graphics},
  19(12):2070--2079, 2013.

\bibitem{Shannon1948}
Claude~E. Shannon.
\newblock A mathematical theory of communication.
\newblock {\em The Bell System Technical Journal}, 27:379--423, 623--656, 1948.

\bibitem{shekhtman2016recent}
Louis~M Shekhtman, Michael~M Danziger, and Shlomo Havlin.
\newblock Recent advances on failure and recovery in networks of networks.
\newblock {\em Chaos, Solitons \& Fractals}, 90:28--36, 2016.

\bibitem{shen2006}
Zeqian Shen, Kwan-Liu Ma, and T.~Eliassi-Rad.
\newblock Visual analysis of large heterogeneous social networks by semantic
  and structural abstraction.
\newblock {\em IEEE Transactions on Visualization and Computer Graphics},
  12(6):1427--1439, Nov 2006.

\bibitem{shi2014Hierarchical}
L.~Shi, Q.~Liao, H.~Tong, Y.~Hu, Y.~Zhao, and C.~Lin.
\newblock Hierarchical focus+context heterogeneous network visualization.
\newblock In {\em IEEE Pacific Visualization Symposium}, pages 89--96,
  \BPrevised{Yokohama, Japan}, March 2014.

\bibitem{Sluban2016Temporal}
Borut Sluban, Miha Gr{\v{c}}ar, and Igor ozeti{\v{c}}.
\newblock Temporal multi-layer network construction from major news events.
\newblock In Hocine Cherifi, Bruno Gon{\c{c}}alves, Ronaldo Menezes, and
  Roberta Sinatra, editors, {\em Complex Networks VII: Proceedings of the
  $7^{th}$ Workshop on Complex Networks CompleNet}, pages 29--41,
  \BPrevised{Dijon, France}, 2016. Springer International Publishing.

\bibitem{SLUSARCZYK201795}
Grażyna \'{S}lusarczyk, Andrzej \L{}achwa, Wojciech Palacz, Barbara Strug,
  Anna Paszy\'{n}ska, and Ewa Grabska.
\newblock An extended hierarchical graph-based building model for design and
  engineering problems.
\newblock {\em Automation in Construction}, 74:95 -- 102, 2017.

\bibitem{smith2006facetmap}
Greg Smith, Mary Czerwinski, B~Robbins Meyers, G~Robertson, and Daniel~Stanley
  Tan.
\newblock Facetmap: A scalable search and browse visualization.
\newblock {\em IEEE Transactions on visualization and computer graphics},
  12(5):797--804, 2006.

\bibitem{srinivasan2018grpahiti}
A.~Srinivasan, H.~Park, A.~Endert, and R.~C. Basole.
\newblock Graphiti: Interactive specification of attribute-based edges for
  network modeling and visualization.
\newblock {\em IEEE Transactions on Visualization and Computer Graphics},
  24(1):226--235, Jan 2018.

\bibitem{stasko2008Jigsaw}
John Stasko, Carsten Görg, and Zhicheng Liu.
\newblock Jigsaw: Supporting investigative analysis through interactive
  visualization.
\newblock {\em Information Visualization}, 7(2):118--132, 2008.

\bibitem{tu2013}
Y.~Tu and H.~W. Shen.
\newblock Graphcharter: Combining browsing with query to explore large semantic
  graphs.
\newblock In {\em IEEE Pacific Visualization Symposium (PacificVis)}, pages
  49--56, \BPrevised{Sydney, NSW, Australia}, 2013.

\bibitem{van2003using}
Frank Van~Ham.
\newblock Using multilevel call matrices in large software projects.
\newblock In {\em IEEE Symposium on Information Visualization}, pages 227--232,
  \BPrevised{Seattle, WA, USA}, 2003. IEEE.

\bibitem{vanVugt2017letters}
Ingeborg van Vugt.
\newblock Using multi-layered networks to disclose books in the republic of
  letters.
\newblock {\em Journal of Historical Network Research}, 1(1):25--51, Oct. 2017.

\bibitem{Vehlow:2015gk}
Corinna Vehlow, Fabian Beck, P~Auw{\"a}rter, and Daniel Weiskopf.
\newblock Visualizing the evolution of communities in dynamic graphs.
\newblock {\em Comput. Graph. Forum ()}, 34(1):277--288, 2015.

\bibitem{verbrugge1979multiplexity}
Lois~M Verbrugge.
\newblock Multiplexity in adult friendships.
\newblock {\em Social Forces}, 57(4):1286--1309, 1979.

\bibitem{viau2010}
C.~Viau, M.~J. McGuffin, Y.~Chiricota, and I.~Jurisica.
\newblock The flowvizmenu and parallel scatterplot matrix: Hybrid
  multidimensional visualizations for network exploration.
\newblock {\em IEEE Transactions on Visualization and Computer Graphics},
  16(6):1100--1108, 2010.

\bibitem{wang2012epidemics}
Y~Wang and G~Xiao.
\newblock Epidemics spreading in interconnected complex networks.
\newblock {\em Physics Letters A}, 376(42-43):2689--2696, 2012.

\bibitem{Ware2008}
Colin Ware and Peter Mitchell.
\newblock Visualizing graphs in three dimensions.
\newblock {\em \BPrevised{ACM Transactions on Applied Perception (TAP)}},
  5(1):2:1--2:15, January 2008.

\bibitem{Wattenberg:2006js}
Martin Wattenberg.
\newblock Visual exploration of multivariate graphs.
\newblock In {\em \BPrevised{Proceedings of the SIGCHI Conference on Human
  Factors in Computing Systems}}, pages \BPrevised{811--819},
  \BPrevised{Montr\'eal, Qu\'ebec, Canada}, 2006. ACM.

\bibitem{WEHMUTH201650}
Klaus Wehmuth, \'Eric Fleury, and Artur Ziviani.
\newblock On multiaspect graphs.
\newblock {\em Theoretical Computer Science}, 651:50 -- 61, 2016.

\bibitem{wilkinson2009history}
Leland Wilkinson and Michael Friendly.
\newblock The history of the cluster heat map.
\newblock {\em The American Statistician}, 63(2):179--184, 2009.

\bibitem{xia2015networkanalyst}
Jianguo Xia, Erin~E Gill, and Robert~EW Hancock.
\newblock Networkanalyst for statistical, visual and network-based
  meta-analysis of gene expression data.
\newblock {\em Nature protocols}, 10(6):823--44, 2015.

\bibitem{yang2016user}
Hui Yang, Kaizhi Tang, Xiong Liu, Lemin Xiao, Roger Xu, and Soundar Kumara.
\newblock A user-centred approach to information visualisation in nano-health.
\newblock {\em International Journal of Bioinformatics Research and
  Applications}, 12(2):95--115, 2016.

\bibitem{yi2007toward}
Ji~Soo Yi, Youn ah~Kang, and John Stasko.
\newblock Toward a deeper understanding of the role of interaction in
  information visualization.
\newblock {\em IEEE transactions on visualization and computer graphics},
  13(6):1224--1231, 2007.

\bibitem{Zeng:2016ez}
An~Zeng and Stefano Battiston.
\newblock The multiplex network of {EU} lobby organizations.
\newblock {\em PLOS ONE}, 11(10):e0158062, 2016.

\bibitem{zhao2013interactive}
Jian Zhao, Christopher Collins, Fanny Chevalier, and Ravin Balakrishnan.
\newblock Interactive exploration of implicit and explicit relations in faceted
  datasets.
\newblock {\em IEEE Transactions on Visualization and Computer Graphics},
  19(12):2080--2089, 2013.

\end{thebibliography}
